\documentclass[aps,pra,floatfix]{revtex4}

\usepackage{graphicx}
\usepackage[english]{babel}
\usepackage{amsmath}
\usepackage{amssymb}
\usepackage{slashed}
\usepackage{cancel}
\usepackage{txfonts}

\allowdisplaybreaks

\newcommand{\me}{\mathrm{e}}
\newcommand{\mi}{\mathrm{i}}

\newcommand{\dif}{\mathrm{d}}

\newcommand{\bk}{\mathbf{k}}

\newcommand{\bx}{\mathbf{x}}

\begin{document}
	\title{$q$-deformed Fermion in Many-Particle Systems and Its Application to BCS Theory}
\author{Xu-Yang Hou$^{1}$, Xun Huang$^{1}$, Yan He$^{2}$, Hao Guo$^{1\ast}$}
\affiliation{$^1$Department of Physics, Southeast University, Nanjing 211189, China}
\affiliation{$^2$College of Physical Science and Technology, Sichuan University, Chengdu, Sichuan 610064, China}

\begin{abstract}
In recent decades, there have been increasing interests in quantum statistics beyond the standard Fermi-Dirac and Bose-Einstein statistics, such as the fractional statistics, quon statistics, anyon statistics and quantum groups, since they can provide some new insights into the cosmology, nuclear physics and condensed matter. In this paper, we study the many-particle system formed by the $q$-deformed fermions ($q$-fermion), which is realized by deforming the quantum algebra of the anticommutation relations. We investigate from a standard perspective of the finite temperature field theory and try to construct the finite temperature Green's function formalism for the free many-$q$-fermion system, then generalize it to the well known interacting fermionic system, the superconductor, and finally obtain a consistent $q$-deformed BCS ($q$BCS) theory.
At low temperature, this theory predicts a Sarma-like ordered phase, and we call it the $q$-deformed Sarma phase. It also presents a symmetric phase diagram in the parameter space and new thermodynamic relations.
\end{abstract}
\email{guohao.ph@seu.edu.cn}
	\maketitle
\section{Introduction}
About three decades ago, the study of solutions to the Yang-Baxter equation led to the concepts of the quantum groups and the associated algebras\cite{Drinfeld,JimboLMP86,JimboBook,qo0}, which are the deformed versions of the ordinary Lie algebra. The new mathematical structures have been applied to the exactly solvable models\cite{Vega} and two-dimensional conformal field theories\cite{Alvarez}. The quantum algebra, such as su$_q$(2), can be realized by the $q$-deformed bosonic harmonic oscillator\cite{qb1,qb2,qb3}, and has a lot of applications in several areas of physics. For example, it plays an important role in the deformed version of statistics and thermodynamics\cite{qh3,qh1,qh5,qh4,VPJC2,qh8,qh6,qh7,qh9,qo6,qh11,qh2}, as well as in the description of spin chain\cite{sc1,sc2}, two-dimensional string theory\cite{qstring2D} and quantum optics\cite{qo1,qo2,qo3,qo4,qo5}. It was also argued that the $q$-deformed algebra can be used to describe the continuous interpolation between Bose and Fermi statistics\cite{qstatis}, and the fractional statistics\cite{VPJC2}.

A natural generalization of the $q$-boson is the introduction of the $q$-fermion, which is associated with the $q$-deformed superalgebras\cite{sqa1,sqa2}, quantum exceptional algebras\cite{eqa} and some $q$-deformed Lie algebras\cite{qa}.
These oscillators, which obey either $q$-deformed commutation or anti-commutation relations, are often called $q$-oscillators.
Interestingly, the anyons also formally satisfy a $q$-deformed commutation relation with $q$ being a complex number of modulus 1 which denotes the anyonic phase shift\cite{qAnyon93,qAnyon96}. However, despite many analogies, there are significant differences between the $q$-oscillators and the anyons. The former are local
operators and can be defined in any dimensions, while the latter are intrinsically non-local and live in strictly two dimensional space.
Nevertheless, since the anyon gas can be realized in the atomic lattice, we may still expect that the $q$-oscillators can be modulated by the anyonic system at least in the two dimensional situation in the future.

There have been many investigations devoted to the studies of the $q$-deformed fermionic oscillator and the associated thermostatistics\cite{VPJC1,qh11,qfs1,qfs2,qfs3,qfs4,Algin11,QF1,QF2,qfm04}. In the more complex systems, the $q$-deformed nucleon pairs were also introduced to study the appearance of condensates in nuclear physics\cite{qfnp,qbcs3,qbcs1,qbcs2,qnjl,qnclp01,qnclp03,qnclp03a,qnclp11}. Recently, it was argued that the newly found photons with half-integral spin\cite{hspt} may be recognized as $q$-deformed fermions\cite{hspt17}, and the $q$-deformed statistics was even applied to discuss the possible emissivity of the light fermionic dark matter in the cooling of the supernova SN1987A\cite{qsn17}.

In this paper, we first generalize the temperature Green's function formalism to the noninteracting gas formed by $q$-fermions. It is found that the single particle Green's function satisfies a new periodicity condition such that the corresponding Matsubara frequency obtains an extra imaginary part relating to the deformed parameter $q$. With the help of this technique, we continue to study the more complex system in condensed matter physics, the superconductor, and construct the $q$-deformed BCS ($q$BCS) theory. Our theoretical model is different from the models building by $q$-deformed nucleon pairs in nuclear physics\cite{qbcs3,qbcs1,qbcs2}, of which the generators of nucleon pairs are supposed to satisfy the su$_q$(2) algebra. This formalism naturally reduces to the model of the noninteracting $q$-Fermi gas when the particle-particle interaction approaches zero, and also reduces to the ordinary BCS theory if $q=1$. Our model has some interesting and special properties. For example, the fermions with different spins must have different but symmetric deformation parameters to ensure the equations of motion for the Green's function and anomalous Green's function form a closed set of coupled equations. Hence the $q$BCS theory naturally has a population-imbalanced structure and present a Sarma-like\cite{Sarma} ordered phase at low temperature. However, it differs from the ordinary Sarma phase in some aspects, such the property of
the number distribution of each species. Even in the population-balanced situation, the system still exhibits the breached pair state\cite{WilczekPRL03} in the momentum space. The theory also presents an interesting phase diagram in the parameter space, and some of the thermodynamic relations are changed due to the emerging of the deformation parameter.

The rest of the paper is organized as follows. In Sec.\ref{sec1}, we briefly review recent studies of the $q$-deformed fermions, and then build the finite temperature Green's function formalism of the noninteracting $q$-deformed Fermi gas. We also study the thermodynamics of it. In Sec.\ref{sec2}, we generalize the study to the ordered system and build the $q$BCS theory. We then give a detailed study on its properties and thermodynamics. The
conclusion is summarized in the end.

\section{Temperature Green's Function Formalism for Noninteracting $q$-Fermions}\label{sec1}
\subsection{The $q$-deformed Fermion}
Throughout this paper, we take the convention that $\hbar=1$, $k_B=1$. For simplicity, we first briefly review the $q$-deformed Fermi algebra and discuss the finite temperature field theoretic formalism for the noninteracting $q$-deformed Fermi gas, then generalize it to the ordered system.

There are four types of known fermionic $q$-deformed algebras\cite{Algin11}, called fermionic Newton (FN), Chaichian-Kulish-Ng (CKN), Parthasarathy-Viswanathan-Chaichian (PVC) and Viswanathan-Parthasarathy-Jagannathan-Chaichian (VPJC) models. Although the algebras of them are different from each other, they actually can be related by certain transformations. For example, the one-dimensional FN-oscillators in fact satisfy the CKN algebra. The most commonly accepted one-dimensional CKN algebra with symmetric deformed parameters is found to be strictly isomorphic to the ordinary one-dimensional fermionic algebra after rescaling the operators\cite{CKN11}. The VPJC-oscillators can be obtained by performing a rescaling transformation on the PVC-oscillators\cite{VPJC1,VPJC2,QF0}.

Here we focus on the VPJC algebra, and then apply it to the noninteracting single-component Fermi gas. The algebra is defined by the following relations
\begin{align}
&\psi_{\bk}\psi^\dagger_{\bk'}+q\psi^\dagger_{\bk'}\psi_{\bk}=\delta_{\bk\bk'},\label{a1}\\
&\left[N, \psi_{\bk}\right]=-\psi_{\bk},\quad
\left[N, \psi^\dagger_{\bk}\right]=\psi^\dagger_{\bk} .\label{a2}
\end{align}
Here $\psi^{(\dagger)}$ is the (creation) annihilation operator for the $q$-fermion, and $N\equiv\sum_\mathbf{k} N_\mathbf{k}$, where $N_\mathbf{k}$ is the fermion number operator with momentum $\mathbf{k}$, is the $q$-deformed total number operator with $q\in R^+$. Note that $\psi$ does not necessarily satisfy $(\psi)^2=0$, and $N_\mathbf{k}\neq \psi^\dagger_\mathbf{k}\psi_\mathbf{k}$ but it nevertheless satisfies the usual relations with $\psi^\dagger_\mathbf{k}$ and $\psi_\mathbf{k}$ as indicated by Eq.(\ref{a2}). The noninteracting Hamiltonian is taken as
\begin{align}\label{nH}
H=\sum_\mathbf{k} \xi_\mathbf{k} N_\mathbf{k},
\end{align}
where $\xi_\mathbf{k}=\epsilon_\bk-\mu=k^2/(2m)-\mu$, with $m$ and $\mu$ being the mass and chemical potential of the $q$-fermion respectively, is the energy dispersion.
When $q=1$, it obviously reduces to the ordinary fermionic oscillator. As pointed previously, this model can be constructed from the PVC-oscillators by a rescaling transformation. 

To discuss the Fock space representation of this model, we introduce the deformed fermion occupation operator $\hat {N}=\psi^\dagger\psi$, where we have omitted the subscript ``$\mathbf{k}$'' for convenience. Let $\{|n\rangle\}$ be the set of basis of the Fock space such that $\hat{N}|n\rangle=g_n|n\rangle$. It can be shown\cite{Algin11} that
\begin{align}\label{gn}
g_n=\frac{1-(-1)^nq^n}{1+q}, \quad |n\rangle=\frac{(\psi^\dagger)^n}{\sqrt{g_n!}}|0\rangle,
\end{align}
where $g_n!=g_ng_{n-1}g_{n-2}\cdots g_1$.
The Pauli exclusion principle is recovered in the limit $q=1$ since we have $\lim_{q\rightarrow1}g_1=1$ and $\lim_{q\rightarrow1}g_2=0$, the latter of which implies $(\psi^\dagger)^2=0$. The non-negative norm condition on the state vector ($g_n\ge 0$) requires $q>0$.

The statistical distribution function of this $q$-gas can be obtained by calculating the ensemble average of the deformed fermion number operator, i.e. $\langle \hat{N}_\mathbf{k}\rangle$. From the commutative relation (\ref{a1}) we can deduce that for any function $F(N_\mathbf{k})$, one has the relation
$\psi_\mathbf{k}F(N_\mathbf{k})=F(N_\mathbf{k}+1)\psi_\mathbf{k}$. Using this relation and the cyclic property of the trace we get
\begin{align}
	n_\mathbf{k}
=\frac{1}{Z}\textrm{Tr}\big( \me^{-\beta H} \psi^\dagger_{\mathbf{k}} \psi_{\mathbf{k}}\big)=\frac{1}{Z}\textrm{Tr}\big( \me^{-\beta \xi_\mathbf{k}(N_\mathbf{k}+1)} \psi_{\mathbf{k}} \psi^\dagger_{\mathbf{k}}\big)=\me^{-\beta \xi_\mathbf{k}}-\me^{-\beta\xi_\mathbf{k}}qn_\mathbf{k},
    \end{align}
    where $n_\bk$ is the number distribution function of the deformed fermions (please notice the difference between $N_\bk$, $\hat{N}_\bk$ and $n_\bk$) and $\beta=\frac{1}{T}$ is the inverse temperature and\begin{eqnarray}\label{gpf}Z\equiv\textrm{Tr}\big( \me^{-\beta H}\big)\end{eqnarray} is the grand partition function. This relation leads to the number distribution of the noninteracting $q$-fermion
\begin{eqnarray}\label{ndf}
n_\mathbf{k}=\frac{1}{q}\frac{1}{1+\frac{1}{q}\me^{\beta \xi_\mathbf{k}}}=\frac{1}{\me^{\beta \xi_\mathbf{k}}+q},
\end{eqnarray}
and the expression of the total particle number
\begin{eqnarray}\label{tpn}
\bar{N}=\sum_\bk\frac{1}{\me^{\beta \xi_\mathbf{k}}+q}.
\end{eqnarray}

\subsection{Thermodynamics}
There have been some discussions about the thermal statistics of various fermionic $q$-oscillators\cite{Algin11,qfs2,VPJC2,qstatis}, all of which are focused on the noninteracting systems. Here we give a systematic discussion about the thermodynamics of such systems via the partition function. Since all thermal quantities can be obtained from it, we need to find a way to evaluate it from the definition (\ref{gpf}). The $q$-deformed number distribution function (\ref{ndf}) can be reexpressed as
  \begin{align}\label{ndf2}
n_\mathbf{k}=\frac{1}{q}\frac{1}{1+\me^{\beta (\xi_\mathbf{k}-\frac{\ln q}{\beta})}},
\end{align}
which can be formally recognized as the number distribution function (normalized by $\frac{1}{q}$ since the fermion is deformed by $q$) of non-deformed fermions if the particle energy is shifted by a temperature-dependent chemical potential $\frac{\ln q}{\beta}$.
 Hence it is reasonable to evaluate the trace in Eq.(\ref{gpf}) over the Fock space of normal fermions with energy shifted by $\frac{\ln q}{\beta}$
\begin{align}
Z&=\sum_n\langle n|\me^{-\beta \sum_\bk(\xi_\bk-\frac{\ln q}{\beta}) N_\bk}|n\rangle\notag\\
&=\sum_{\sum_\bk n_\bk=n;n_\bk=0,1}\prod_\mathbf{k}\otimes\langle n_\bk |\me^{-\beta \sum_\bk(\xi_\bk-\frac{\ln q}{\beta}) N_\bk}\prod_\mathbf{k}\otimes| n_\bk\rangle\notag\\
&=\prod_\mathbf{k}\sum_{n_\bk=0,1}\me^{-\beta \sum_\bk(\xi_\bk-\frac{\ln q}{\beta}) n_\bk}\notag\\
&=\prod_\mathbf{k}(1+qz\me^{-\beta\epsilon_\mathbf{k}}),
\end{align}
where $z=\me^{\beta\mu}$ is the fugacity. For $q$-oscillator systems, some of the standard thermodynamical relations in the usual form are ruled out, for instance $\bar{N}\neq z\frac{\partial \ln Z}{\partial z}$ \cite{qh2}. It in fact needs a renormalization since the number distribution function is normalized by $1/q$ according to Eq.(\ref{ndf2})
 \begin{align}\label{qN}
 \bar{N}=\frac{1}{q}z\frac{\partial \ln Z}{\partial z}=\sum_\mathbf{k}\frac{1}{\me^{\beta \xi_\mathbf{k}}+q}.
 \end{align}
 The inclusion of the normalization factor $\frac{1}{q}$ is consistent with our later discussions about the Green's function formalism. In the standard manner, the thermodynamic potential is given by
 \begin{align}\label{qO}
\Omega=-\frac{1}{\beta}\ln Z=-\frac{1}{\beta}\sum_\mathbf{k}\ln(1+q\me^{-\beta\xi_\mathbf{k}}).
 \end{align}
 Hence the number density equation (\ref{qN}) can be reexpressed as
  \begin{align}\label{qN1}
 \bar{N}=-\frac{1}{q}\frac{\partial \Omega}{\partial \mu}.
 \end{align}
The pressure can be determined in the thermodynamic limit, which leads to the equation of state
 \begin{align}
 PV=-\Omega=k_BT\sum_\mathbf{k}\ln(1+q\me^{-\beta\xi_\mathbf{k}}).
 \end{align}
 To evaluate to the total energy of the system, we also need to include the normalization factor
  \begin{align}\label{qE}
  E=-\frac{1}{q}\frac{\partial \ln Z}{\partial \beta}+\mu\bar{N}=\sum_\mathbf{k}\frac{\epsilon_\mathbf{k}}{\me^{\beta \xi_\mathbf{k}}+q},
  \end{align}
The entropy is given by the traditional procedure
  \begin{align}\label{qS}
 S=-\frac{\partial \Omega}{\partial T}=-\sum_\bk\Big[ f(\xi_\bk-\frac{\ln q}{\beta})\ln\frac{1}{q}f(\xi_\bk-\frac{\ln q}{\beta})+
f(-\xi_\bk+\frac{\ln q}{\beta})\ln f(-\xi_\bk+\frac{\ln q}{\beta})\Big].
\end{align}
where $f(x)=1/(\me^{x/T}+1)$ is the Fermi distributive function. Since the $q$-fermion system is nonextensive, the entropy is non-additive\cite{NonExSyS,NonExbook}. By applying Eqs.(\ref{qN}), (\ref{qO}), (\ref{qE}) and (\ref{qS}), it is straightforward to verify the thermodynamic identity
  \begin{align}\label{qEOSN}
  E=\frac{1}{q}(\Omega+TS)+\mu \bar{N}.
  \end{align}
  Obviously when $q=1$ it reduces to the well-known identity for the un-deformed noninteracting Fermi gas.

\subsection{Finite temperature Green's function formalism}
Now we turn to the finite temperature formalism for the $q$-deformed many-fermion system by introducing the imaginary time $\tau=\mi t$. Define the 4-coordinate $x$ as $x=(\tau,\bx)$. The finite temperature Heisenberg operator is obtained by the transformation $\mathcal{O}(x)=\me^{H\tau}\mathcal{O}(\bx)\me^{-H\tau}$.
	The fermionic $q$-deformed single-particle Green's function is defined by
	\begin{align}
	G(x,x')=-\langle T_\tau[\psi(x)\psi^\dagger (x')]\rangle=-\langle\psi(x)\psi^\dagger(x')\rangle\theta(\tau-\tau')+q\langle\psi^\dagger(x')\psi(x)\rangle\theta(\tau'-\tau).
	\end{align}
This Green's function has a different periodicity property from that of the ordinary fermions. If the system has a spacetime translational symmetry, for $0<\tau<\beta$ one can show
\begin{align}
G(-\tau,\bx-\bx')=-qG(-\tau+\beta,\bx-\bx').\label{p10}
\end{align}
Details can be found in Appendix.\ref{appa1}. Similarly
\begin{align}
G(\tau-\beta,\bx-\bx')=-qG(\tau,\bx-\bx').\label{p2}
\end{align}
In general we have
\begin{align}
G(-\beta<\tau-\tau'<\beta,\bx-\bx')=-qG(\tau-\tau'+\beta,\bx-\bx').\label{p3}
\end{align}
The definition of the Green's function can be generalized to the whole region $-\infty<\tau-\tau'<+\infty$ by
\begin{align}
G(\tau-\tau',\bx-\bx')=G(\tau-\tau'+2\beta,\bx-\bx').\end{align}
 The associated Fourier transformation of the Green's function is given by \begin{align}\label{FTG}G(\omega_n,\bx)=\frac{1}{2}\int^{\beta}_{-\beta}d\tau \me^{\mi\omega_n\tau}G(\tau,\bx).\end{align}
The periodicity property of the Green's function leads to the fact that the Matsubara frequency has an extra imaginary part
\begin{eqnarray}\label{mf}
	\omega_n=\frac{(2n+1)\pi}{\beta}-\mi\frac{\ln q}{\beta}.
	\end{eqnarray}
For details, please refer to Appendix.\ref{appa1}.
Similar result for $q$-bosons was obtained in Ref.\cite{VPJC2}, where the bosonic Matsubara frequency is
\begin{eqnarray}\label{mb}
	\omega_n=\frac{2n\pi}{\beta}-\mi\frac{\ln q}{\beta}.
	\end{eqnarray}
Although the deformation parameter $q$ is restricted as $q\in R^+$, the non-negative condition (\ref{gn}) only requires $q>-1$. In fact, the $q$-fermion can be extrapolated to the $q$-bosons if $q$ can be deformed from positive to negative. This can be clarified by noting that the algebra (\ref{a2}) becomes
\begin{eqnarray}
\psi_{\bk}\psi^\dagger_{\bk'}-|q|\psi^\dagger_{\bk'}\psi_{\bk}=\delta_{\bk\bk'}
\end{eqnarray}
which is the commutation relation of the $q$-bosons \cite{VPJC2}. Naively, if the $q$ is assumed to be negative in the expression (\ref{mf}), the Matsubara frequency for $q$-fermions becomes
\begin{align}\label{mfb}
	\omega_n=\frac{(2n+1)\pi}{\beta}-\mi\frac{\ln |q|+\mi\arg(-1)}{\beta}=\frac{(2n+2)\pi}{\beta}-\mi\frac{\ln |q|}{\beta},
	\end{align}
where $\arg(-1)=\pi$ is the argument of $q$.
Interestingly it accordingly becomes the Matsubara frequency for $q$-bosons. Therefore, as the deformation parameter evolves from positive to negative except the singular point $q=0$, the $q$-Fermi gas can be connected to the $q$-Bose gas by naively changing the sign of $q$.

Given the definition of the Heisenberg operator $\psi(x)=\me^{H\tau}\psi(\bx)\me^{-H\tau}$, we have the equation of motion for the field operator
\begin{eqnarray}
	\frac{\partial \psi(x)}{\partial \tau}=-\big(\frac{-\nabla^2}{2m}-\mu\big)\psi(x).
	\end{eqnarray}
By applying this relation we obtain the equation of motion for the Green's function
	\begin{eqnarray}\label{eom}
	\left(-\frac{\partial }{\partial \tau}-\frac{-\nabla^2}{2m}+\mu\right)G(x,x')&=&\delta(\bx-\bx')\delta(\tau-\tau').
	\end{eqnarray}
	Again, details can be found in Appendix.\ref{appa1}. This equation can be easily solved by performing the Fourier transformation
\begin{eqnarray}\label{G2}
G(x-x')
=\sum_K\me^{\mi\mathbf{k}\cdot(\mathbf{x}-\mathbf{x}')-\mi\omega_n(\tau-\tau')}G(\mi\omega_n,\mathbf{k}),
\end{eqnarray}
where the Matsubara frequency is given by Eq.(\ref{mf}), $K=(\mi\omega_n,\mathbf{k})$ and $\sum_K\equiv T\sum_{\mi\omega_n}\sum_{\mathbf{k}}$.
	In momentum space, the equations (\ref{eom}) become
	\begin{eqnarray}
	(\mi\omega_n-\xi_\bk)G(\mi\omega_n,\bk)=1.
	\end{eqnarray}
 The solution of this equation is
	\begin{eqnarray}\label{G0}
	G(\mi\omega_n,\bk)=\frac{1}{\mi\omega_n-\xi_\mathbf{k}}=\frac{1}{\mi\frac{(2n+1)\pi}{\beta}-(\xi_\mathbf{k}-\frac{1}{\beta}\ln q)}.
		\end{eqnarray}
Again, we can see that $\frac{1}{\beta}\ln q$ can be recognized as the effective temperature-dependent chemical potential.
To avoid further confusion, we will still use $\mi\omega_n=\mi\frac{(2n+1)\pi}{\beta}$ as the Matsubara frequency hereafter, and $G(\mi\omega_n,\bk)$ is reinterpreted as $G(\mi\frac{(2n+1)\pi}{\beta},\bk)$ instead of $G(\mi\frac{(2n+1)\pi}{\beta}+\frac{1}{\beta}\ln q,\bk)$ for later convenience.
The number distribution function 
can also be obtained from the Green's function. Let $x^+=(\tau^+,\bx)$,the particle number of the $q$-gas is given by
    \begin{eqnarray}
    N=\int d^3\bx\langle \psi^\dagger(\bx)\psi(\bx)\rangle=\frac{1}{q}\int d^3\bx G(x,x^+)=\frac{V}{q}\sum_\bk f(\xi_\bk-\frac{1}{\beta}\ln q),
    \end{eqnarray}
    where the Matsubara frequency summation has been implemented. The normalization factor $\frac{1}{q}$ also appears here, which is consistent with the previous discussions in thermodynamics.
     Therefore we get
    \begin{eqnarray}
    n=\frac{N}{V}=\sum_\bk\frac{1}{\me^{\beta\xi_\bk}+q}.
    \end{eqnarray}
    Hence the number density with momentum $\mathbf{k}$ is
    $    n_\bk=\frac{1}{\me^{\beta \xi_\bk}+q}$
    which is exactly the same as Eq.(\ref{ndf}). The ground state of the $q$-deformed Fermi gas is also a Fermi sea normalized by $\frac{1}{q}$, which can be deduced by noting that $n_\bk=\frac{1}{q}\theta(k-k_F)$ at zero temperature with $k_F=\sqrt{2m\mu}$ being the Fermi momentum. The relation between the number density and the Fermi momentum can also be obtained by

    \begin{eqnarray}\label{nfm}
    n=\frac{1}{q} \int \frac{d^3k}{(2\pi)^3} \theta(k_F-k)=\frac{k_F^3}{6q\pi^2}.
    \end{eqnarray}
    Therefore $k_F=\sqrt[3]{6q\pi^2n}$.

\section{$Q$-deformed BCS Theory}\label{sec2}
In this section, we still use the same notations like $H$, $\bar{N}$, $\cdots$, etc. to denote important physical quantities such as the Hamiltonian, partial number, $\cdots$, etc.. Please notice the difference of their meanings.
\subsection{Basic Formalism}
Now we generalize our discussion to the interacting fermionic system. One of the most famous examples is the superconducting system which is well understood by the BCS theory with the Hamiltonian
\begin{equation}
H=\sum_{\bk}(\sum_\sigma \xi_\bk \psi^\dagger_{\bk\sigma}\psi_{\bk\sigma}+\Delta
^\ast \psi_{-\bk\uparrow} \psi_{\bk\downarrow}+\Delta\psi^\dagger_{\bk\downarrow}\psi^\dagger_{-\bk\uparrow})+\frac{|\Delta|^2}{g},
\end{equation}
where $\psi^{(\dagger)}_{\mathbf{k}\sigma}$ with $\sigma=\uparrow,\downarrow$ is the annilation (creation) operator for the $q$-deformed fermion with different spin, $\Delta(\mathbf{x})=g\langle\psi_\uparrow(\mathbf{x})\psi_\downarrow(\mathbf{x})\rangle$ is the order parameter or the pairing gap fucntion, and $g$ is the attractive coupling constant. By defining \begin{align}\label{NSS}N_{\bk \sigma}=\psi^\dagger_{\bk\sigma}\psi_{\bk\sigma}, S_\bk=\psi_{-\bk\uparrow} \psi_{\bk\downarrow},S^\dagger_\bk=\psi^\dagger_{\bk\downarrow}\psi^\dagger_{-\bk\uparrow},\end{align} the Hamiltonian can be into the form
\begin{equation}\label{HqBCS0}
H=\sum_{\bk,\sigma}\xi_{\bk\sigma}N_{\bk\sigma}+\sum_\bk(\Delta^* S_\bk+\Delta S^\dag_\bk)+\frac{|\Delta|^2}{g},
\end{equation}
which is a starting point to construct the $q$ analog of the BCS theory. Here $N_{\bk\sigma}$ is the corresponding number operator for each species,
$S_\bk$ and $S^\dagger_\bk$ are spin operators satisfying the algebra
\begin{align}\label{nSa}
&\Big[S^\dag_{\bk},\psi_{\bk'\uparrow}\Big]=\psi^\dag_{-\bk\downarrow}\delta_{\bk\bk'}, \quad \Big[S^\dag_{\bk},\psi_{\bk'\downarrow}\Big]=-\psi^\dag_{-\bk\uparrow}\delta_{\bk\bk'}\notag \\
&\Big[S_{\bk},\psi^\dag_{\bk'\uparrow}\Big]=-\psi_{-\bk\downarrow}\delta_{\bk\bk'},  \quad \Big[S_{\bk},\psi^\dag_{\bk'\downarrow}\Big]=\psi_{-\bk\uparrow}\delta_{\bk\bk'},\notag\\
&\textrm{all other commutators vanish.}
\end{align}

To construct the $q$BCS theory, instead of the definition (\ref{NSS}) we introduce the number operator and spin operators by generalizing the relations (\ref{a2}) and (\ref{nSa}). In other words, the Hamiltonian of the $q$BCS theory takes the expression
\begin{equation}\label{HqBCS}
H=\sum_{\bk,\sigma}\xi_{\bk\sigma}N_{\bk\sigma}+\sum_\bk(\Delta^* S_\bk+\Delta S^\dag_\bk)+\frac{|\Delta|^2}{qg},
\end{equation}
and the the number operator and spin operators never take the form (\ref{NSS}), they are instead defined by giving the algebraic relations between them and field operators. Here the constant term is normalized by the deformation parameter, and later we will find that this is necessary to give the correct thermodynamic relations. Comparing to the noninteracting $q$-gas, the interaction of the $q$BCS model is introduced via the deformed spin operator.
 Now we explicitly give the relations step by step.
The system has two sets of field operators (with up-spin and down-spin respectively), they are assumed to independently satisfy the following deformed algebra
\begin{align}\label{qBCSa1}
&\psi_{\bk\uparrow}\psi^\dagger_{\bk'\uparrow}+q\psi^\dagger_{\bk'\uparrow}\psi_{\bk\uparrow}=\delta_{\bk\bk'},\quad
\psi_{\bk\downarrow}\psi^\dagger_{\bk'\downarrow}+q^{-1}\psi^\dagger_{\bk'\downarrow}\psi_{\bk\downarrow}=\delta_{\bk\bk'},\notag\\
&\psi_{\bk\uparrow}\psi_{\bk'\downarrow}+q\psi_{\bk'\downarrow}\psi_{\bk\uparrow}=0,\quad \quad 
\psi^\dag_{\bk\downarrow}\psi^\dag_{\bk'\uparrow}+q\psi^\dag_{\bk'\uparrow}\psi^\dag_{\bk\downarrow}=0,
\end{align}
where $\bar{\uparrow}=\downarrow$ and vice versa. This is a two-mode generalization of the VPJC algebra, and the deformation parameter of one species is the inverse of that of the other. The first line indicates that the field operators satisfy $q$ and $q^{-1}$ deformed algebra respectively, which is to ensure the differential equations of the Green's function and anomalous Green's function to form a closed set. This is because that the equation of motion for one species is involved with that of another. The necessarity of the introducing of two deformation parameters will become clear in later discussions of the Green's functions, details can be found in the derivation and discussion of Eq.(\ref{eomu}). The other lines are also designed in order to get a solvable $q$BCS theory. The second line is crucial to the property of the anomalous Green's function and is self-consistent if the spin is flipped. This algebra can be constructed by choosing suitable set of parameters for the multi-parameter deformed fermionic oscillators presented in Ref.\cite{cqf1}. The number operator for each species satisfy the following algebra
\begin{align}\label{qNa}
&\big[N_{\sigma},\psi_{\bk\sigma}\big]=-\psi_{\bk\sigma},\quad
\big[N_{\sigma},\psi^\dagger_{\bk\sigma}\big]=\psi^\dagger_{\bk\sigma},\quad
\big[N_{\sigma},\psi_{\bk\bar{\sigma}}\big]=0,
\end{align}
where $N_\sigma=\sum_\bk n_{\bk \sigma}$, and this is a direct generalization of Eq.(\ref{a2}). Similarly, by generalizing Eq.(\ref{nSa}), the spin operators satisfy the relation
\begin{align}\label{qSa}
&\Big[S^\dag_{\bk},\psi_{\bk'\uparrow}\Big]=\frac{1}{q}\psi^\dag_{-\bk\downarrow}\delta_{\bk\bk'},  \quad \Big[S^\dag_{\bk},\psi_{\bk'\downarrow}\Big]=-\psi^\dag_{-\bk\uparrow}\delta_{\bk\bk'}\notag \\
&\Big[S_{\bk},\psi^\dag_{\bk'\uparrow}\Big]=-\frac{1}{q}\psi_{-\bk\downarrow}\delta_{\bk\bk'},  \quad \Big[S_{\bk},\psi^\dag_{\bk'\downarrow}\Big]=\psi_{-\bk\uparrow}\delta_{\bk\bk'},\notag\\
&\textrm{all other commutators vanish.}
\end{align}
These operators can also be explicitly constructed by multi-parameter deformed fermionic oscillators if the number of the parameters is large enough. Applying these algebras, it can be found that the Hamiltonian and the field operator satisfy the relation
\begin{align} \label{qBCSa2}
&\Big[H,\psi_{\bk\uparrow}\Big]=-\xi_{\bk\uparrow}\psi_{\bk\uparrow}+q^{-1}\Delta\psi^\dag_{-\bk\downarrow},  \quad \Big[H,\psi_{\bk\downarrow}\Big]=-\xi_{\bk\downarrow}\psi_{\bk\downarrow}-\Delta\psi^\dag_{-\bk\uparrow}, \notag \\
&\Big[H,\psi^\dag_{\bk\uparrow}\Big]=\xi_{\bk\uparrow}\psi^\dag_{\bk\uparrow}-q^{-1}\Delta^* \psi_{-\bk\downarrow},  \quad \Big[H,\psi^\dag_{\bk\downarrow}\Big]=\xi_{\bk\downarrow}\psi^\dag_{\bk\downarrow}+\Delta^* \psi_{-\bk\uparrow}.
\end{align}
The Heisenberg operator is defined as before
$\psi_\sigma(x)=\me^{H\tau}\psi_\sigma(\bx)\me^{-H\tau}$, $\psi^\dag_\sigma(x)=\me^{H\tau}\psi^\dag_\sigma(\bx)\me^{-H\tau}$
where $H$ is the $q$BCS Hamiltonian given by Eq.(\ref{HqBCS}).
By applying this algebra (\ref{qBCSa2}), we can get the equations of motion for the field operator
\begin{align}\label{eomf}
\frac{\partial \psi_\uparrow(x)}{\partial \tau}&=-\left(\frac{(-\mi\nabla)^2}{2m}-\mu_\uparrow \right)\psi_\uparrow(x)+\frac{\Delta(\bx)}{q}\psi^\dagger_\downarrow(x),\notag\\
\frac{\partial \psi^\dagger_\downarrow(x)}{\partial \tau}&=\left(\frac{(-\mi\nabla)^2}{2m}-\mu_\downarrow\right)\psi^\dagger_{\downarrow}(x)+\Delta^\ast(\bx)\psi_{\uparrow}(x).
\end{align}

The single-particle Green's function and anomalous Green's function are defined by
\begin{align}\label{GAG}
&G_{\uparrow}(x,x')=-\langle T_{\tau}[\psi_{\uparrow}(x)\psi^{\dag}_{\uparrow}(x')]\rangle=-\langle\psi_{\uparrow}(x)\psi^{\dag}_{\uparrow}(x')\rangle\theta(\tau-\tau')+q\langle\psi^{\dag}_{\uparrow}(x')\psi_{\uparrow}(x)\rangle\theta(\tau'-\tau), \nonumber \\
&G_{\downarrow}(x,x')=-\langle T_{\tau}[\psi_{\downarrow}(x)\psi^{\dag}_{\downarrow}(x')]\rangle=-\langle\psi_{\downarrow}(x)\psi^{\dag}_{\downarrow}(x')\rangle\theta(\tau-\tau')+q^{-1}\langle\psi^{\dag}_{\downarrow}(x')\psi_{\downarrow}(x)\rangle\theta(\tau'-\tau), \nonumber \\
&F_{\uparrow\downarrow}(x,x')=-\langle T_{\tau}[\psi_{\uparrow}(x)\psi_{\downarrow}(x')]\rangle=-\langle\psi_{\uparrow}(x)\psi_{\downarrow}(x')\rangle\theta(\tau-\tau')+q\langle\psi_{\downarrow}(x')\psi_{\uparrow}(x)\rangle\theta(\tau'-\tau), \nonumber \\
&F_{\downarrow\uparrow}(x,x')=-\langle T_{\tau}[\psi_{\downarrow}(x)\psi_{\uparrow}(x')]\rangle=-\langle\psi_{\downarrow}(x)\psi_{\uparrow}(x')\rangle\theta(\tau-\tau')+q^{-1}\langle\psi_{\uparrow}(x')\psi_{\downarrow}(x)\rangle\theta(\tau'-\tau), \nonumber \\
\end{align}
The definitions are consistent with the algebras (\ref{qBCSa1}) satisfied by the field operators.
Since the deformation parameters for each species are different, we must define the Green's functions regarding to $q$ and $\frac{1}{q}$ respectively. Accordingly, $G_\uparrow$ and $G_\downarrow$ have different periodicity properties
\begin{align}\label{pBCS}
G_{\uparrow}(-\tau,\bx-\bx')&=-qG_{\uparrow}(-\tau+\beta,\bx-\bx'),\notag  \\
G_{\downarrow}(-\tau,\bx-\bx')&=-q^{-1}G_{\downarrow}(-\tau+\beta,\bx-\bx'),
\end{align}
 and so do $F_{\uparrow\downarrow}$ and $F_{\downarrow\uparrow}$.
 Applying the equations of motion (\ref{eomf}) for the field operators, we can derive the equations of motion for the Green's functions
\begin{align}\label{eom1}
&(-\frac{\partial}{\partial \tau}-\frac{(-\mi\nabla)^2}{2m}+\mu_{\uparrow})G_{\uparrow}(x,x')+q^{-1}\Delta(\bx)F^{\dag}_{\uparrow\downarrow}(x,x')=\delta(\bx-\bx')\delta(\tau-\tau'), \notag \\
&(\frac{\partial}{\partial \tau}-\frac{(-\mi\nabla)^2}{2m}+\mu_{\downarrow})F^{\dag}_{\uparrow\downarrow}(x,x')=\Delta^{*}(\bx)G_{\uparrow}(x,x'),
\end{align}
and
\begin{align}\label{eom2}
&(-\frac{\partial}{\partial \tau}-\frac{(-\mi\nabla)^2}{2m}+\mu_{\downarrow})G_{\downarrow}(x,x')-\Delta(\bx)F^{\dag}_{\downarrow\uparrow}(x,x')=\delta(\bx-\bx')\delta(\tau-\tau'), \notag \\
&(\frac{\partial}{\partial \tau}-\frac{(-\mi\nabla)^2}{2m}+\mu_{\uparrow})F^{\dag}_{\downarrow\uparrow}(x,x')=-q^{-1}\Delta^{*}(\bx)G_{\downarrow}(x,x'),
\end{align}
Details can be found in Appendix.\ref{appb1}.
We emphasize that the algebras (\ref{qBCSa1}), (\ref{qNa}) and (\ref{qSa}) are very crucial to the closure of the two sets of differential equations. Since $G_\uparrow$ and $F^\dagger_{\uparrow\downarrow}$ appear in the same set of equations, they must have the same periodicity in the momentum space, hence it is reasonable to define them as in Eqs.(\ref{GAG}), so do $G_\downarrow$ and $F^\dagger_{\downarrow\uparrow}$.
Introducing $\mu=\frac{\mu_\uparrow+\mu_\downarrow}{2}$ and $h=\frac{\mu_\uparrow-\mu_\downarrow}{2}$, and implementing the Fourier transformation, the coupled equations become
\begin{align}\label{dEGAG}
&(\mi\omega_n+\frac{\ln q}{\beta}-\xi_{\bk\uparrow})G_{\uparrow}(\mi\omega_n,\bk)+\frac{1}{q}\Delta F^{\dag}_{\uparrow\downarrow}(\mi\omega_n,\bk)=1,\notag  \\
&(-\mi\omega_n-\frac{\ln q}{\beta}-\xi_{\bk\downarrow})F^{\dag}_{\uparrow\downarrow}(\mi\omega_n,\bk)-\Delta^{*}G_{\uparrow}(\mi\omega_n,\bk)=0.
\end{align}
and
\begin{align}\label{mEGAG}
&(\mi\omega_n-\frac{\ln q}{\beta}-\xi_{\bk\downarrow})G_{\downarrow}(\mi\omega_n,\bk)-\Delta F^{\dag}_{\downarrow\uparrow}(\mi\omega_n,\bk)=1,  \notag\\
&(-\mi\omega_n+\frac{\ln q}{\beta}-\xi_{\bk\uparrow})F^{\dag}_{\downarrow\uparrow}(\mi\omega_n,\bk)+\frac{1}{q}\Delta^{*}G_{\downarrow}(\mi\omega_n,\bk)=0.
\end{align}
The Green's function and anomalous Green's function satisfy two linear equations in the momentum space, then they must have the same periodicity property with respect to the frequency. Hence they must be defined in the same way as shown by Eqs.(\ref{GAG}), which requires that the algebra relation between fermion operators with different spins must be given by the second line of Eqs.(\ref{qBCSa1}). The closure of Eqs.(\ref{eom1}) further restricts the algebra of the spin operator to be Eqs.(\ref{qSa}).

Eqs.(\ref{dEGAG}) and (\ref{mEGAG}) are readily solved to give
\begin{align}\label{eud}
 G_{\uparrow}(\mi\omega_n,\bk)&=\frac{|u_{\bk}|^2}{\mi\omega_n+\frac{\ln q}{\beta}-E_{\bk\uparrow}}+\frac{|v_{\bk}|^2}{\mi\omega_n+\frac{\ln q}{\beta}+E_{\bk\downarrow}},  \notag \\
 F^{\dag}_{\uparrow\downarrow}(\mi\omega_n,\bk)&=-\sqrt{q} u_{\bk}^{*}v_{\bk}\Big(\frac{1}{\mi\omega_n+\frac{\ln q}{\beta}-E_{\bk\uparrow}}-\frac{1}{\mi\omega_n+\frac{\ln q}{\beta}+E_{\bk\downarrow}}\Big),
\end{align}
and
\begin{align}\label{edu}
 G_{\downarrow}(\mi\omega_n,\bk)&=\frac{|u_{\bk}|^2}{\mi\omega_n-\frac{\ln q}{\beta}-E_{\bk\downarrow}}+\frac{|v_{\bk}|^2}{\mi\omega_n-\frac{\ln q}{\beta}+E_{\bk\uparrow}},  \notag\\
\quad \quad F^{\dag}_{\downarrow\uparrow}(\mi\omega_n,\bk)&=\frac{u_{\bk}^{*}v_{\bk}}{\sqrt{q}}\Big(\frac{1}{\mi\omega_n-\frac{\ln q}{\beta}-E_{\bk\downarrow}}-\frac{1}{\mi\omega_n-\frac{\ln q}{\beta}+E_{\bk\uparrow}}\Big),
\end{align}
where $E_{\bk}=\sqrt{\xi_{\bk}^2+\frac{1}{q}|\Delta|^2}$, $\xi_\bk=\frac{\mathbf{k}^2}{2m}-\mu$, $E_{\bk\uparrow,\downarrow}=E_\bk\mp h$ and $|u_{\bk}|^2,|v_{\bk}|^2=\frac{1}{2}(1\pm\frac{\xi_{\bk}}{E_{\bk}})$.
One can see that if $q=1$ the $q$BCS theory reduces to the population-imbalanced BCS theory\cite{Chien06}. At $T=0$ the ground state of our $q$BCS theory is almost the same as that of the ordinary BCS theory except the inclusion of deformation parameters.

Knowing the Green's functions, we can now deduce the self-energy by using the Dyson's relation. Introducing the 4-momentum $K=(\mi\omega_n,\mathbf{k})$ which follows the usual convention, according to Eq.(\ref{G0}) the bare Green's functions for the spin-up fermions deformed by $q$ and spin-down fermions deformed by $\frac{1}{q}$ are respectively given by
\begin{align}
G_{0\uparrow}(K)\equiv \frac{1}{\mi \omega_n+\frac{\ln q}{\beta}-\xi_{\bk\uparrow}},\quad
G_{0\downarrow}(K)\equiv \frac{1}{\mi \omega_n-\frac{\ln q}{\beta}-\xi_{\bk\downarrow}}.
\end{align}
Comparing with the expressions of $G_{\uparrow,\downarrow}(K)$, it is easy to find
\begin{align}
G_{\uparrow}(K)^{-1}=G_{0\uparrow}(K)^{-1}-\Sigma_\uparrow(K),\quad
G_{\downarrow}(K)^{-1}=G_{0\downarrow}(K)^{-1}-\Sigma_\downarrow(K),
\end{align}
where
\begin{align}\label{se}
\Sigma_\uparrow(K)\equiv -\frac{|\Delta|^2}{q}G_{0\downarrow}(-K)=\frac{\frac{1}{q}|\Delta|^2}{\mi\omega_n+\frac{\ln q}{\beta}+\xi_{\bk\downarrow}},\notag\\
\Sigma_\downarrow(K)\equiv -\frac{|\Delta|^2}{q}G_{0\uparrow}(-K)=\frac{\frac{1}{q}|\Delta|^2}{\mi\omega_n-\frac{\ln q}{\beta}+\xi_{\bk\uparrow}}.
\end{align}
 The equations of states for $q$BCS theory can be obtained from the Green's functions. Let $x^+=(\tau^+,\bx)$ with $\tau^+=\tau+0^+$, the particle number densities for spin-up and spin-down fermions can be given respectively
 \begin{align}\label{Number}
n_{\uparrow}&=\frac{1}{V}\int\dif^3\bx\;\langle\psi^{\dag}_{\uparrow}(x)\psi_{\uparrow}(x)\rangle =\frac{1}{qV}\int\dif^3\bx\; G_{\uparrow}(x,x^{+}) = \frac{1}{q}\sum_{\bk}\Big[|u_{\bk}|^2f(E_{\bk\uparrow}-\frac{\ln q}{\beta})+|v_{\bk}|^2f(-E_{\bk\downarrow}-\frac{\ln q}{\beta})\Big], \nonumber \\
n_{\downarrow}&=\frac{1}{V}\int\dif^3\bx\;\langle\psi^{\dag}_{\downarrow}(x)\psi_{\downarrow}(x)\rangle =\frac{q}{V}\int\dif^3\bx\; G_{\downarrow}(x,x^{+}) =q\sum_{\bk}\Big[|u_{\bk}|^2f(E_{\bk\downarrow}+\frac{\ln q}{\beta})+|v_{\bk}|^2f(-E_{\bk\uparrow}+\frac{\ln q}{\beta})\Big].
\end{align}
The quasiparticle energy dispersions are shifted by the deformation parameter at finite temperature and become $E_{\bk\uparrow}-\frac{\ln q}{\beta}$, and $E_{\bk\downarrow}+\frac{\ln q}{\beta}$.
Similarly, the gap equation can be obtained by either one of the anomalous Green's functions
\begin{align}
\Delta^{*}=g\langle\psi^{\dag}_{\downarrow}(\bx)\psi^{\dag}_{\uparrow}(\bx)\rangle=-gF^{\dag}_{\uparrow\downarrow}(x^{+},x) =gqF^{\dag}_{\downarrow\uparrow}(x^{+},x)=g\sum_{\bk}\frac{\Delta^{*}}{2E_{\bk}}\Big[f(E_{\bk\uparrow}-\frac{\ln q}{\beta})-f(-E_{\bk\downarrow}-\frac{\ln q}{\beta})\Big],
\end{align}
which implies
\begin{eqnarray}\label{geq}
-\frac{1}{g}=\sum_{\bk}\frac{1-f(E_{\bk\uparrow}-\frac{\ln q}{\beta})-f(E_{\bk\downarrow}+\frac{\ln q}{\beta})}{2E_\bk}.
\end{eqnarray}
Details can be found in Appendix.\ref{appb1}.

Comparing to the $q=1$ population-imbalanced BCS model\cite{Chien06}, the expression of $n_{\uparrow,\downarrow}$ must include the corresponding deformation parameters $q$ and $\frac{1}{q}$ respectively just as the non-interacting situation, the order parameter is normalized as $\Delta\rightarrow\frac{\Delta}{\sqrt{q}}$, and $h$ is shifted as $h\rightarrow h+T\ln q$.

\begin{figure}[ht]
\centering
\includegraphics[width=2.4in, clip]{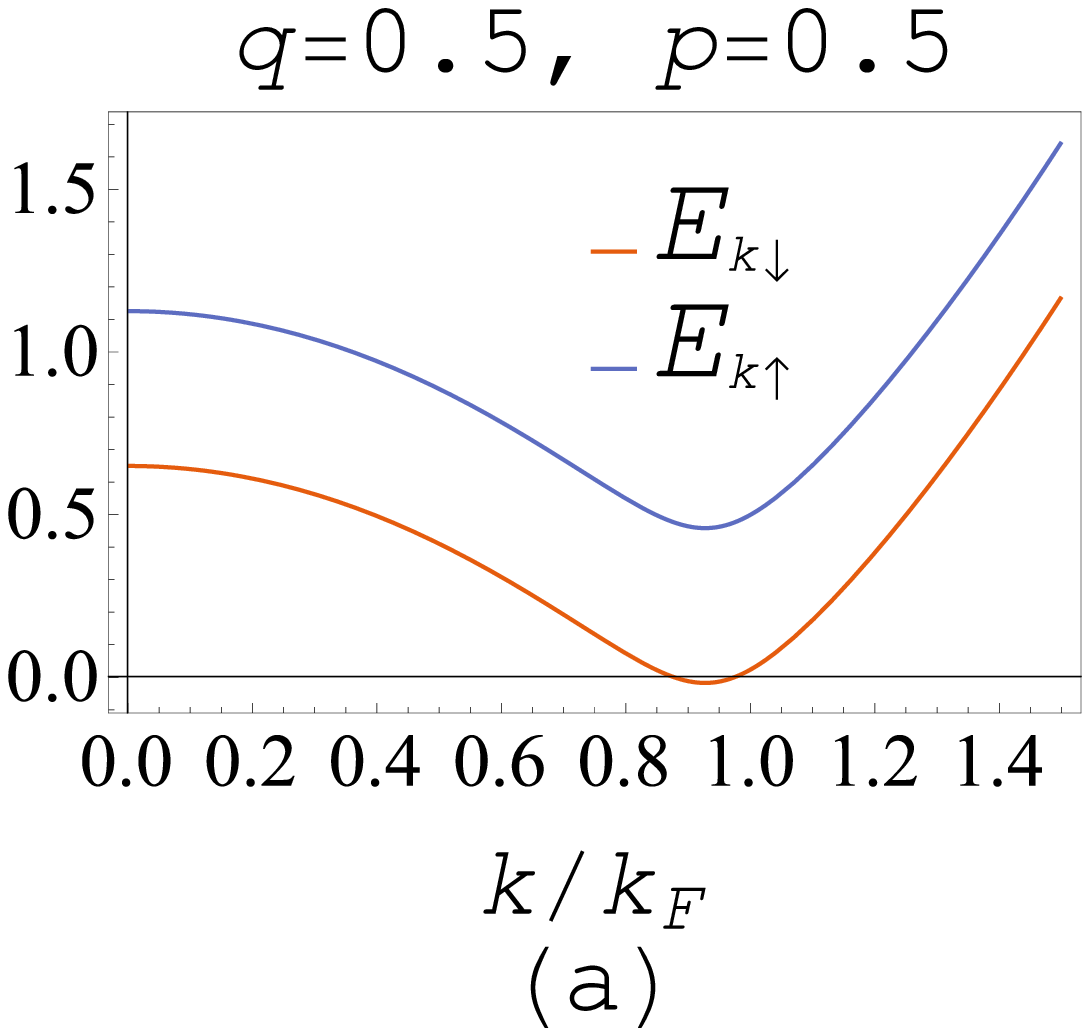}
\includegraphics[width=2.4in, clip]{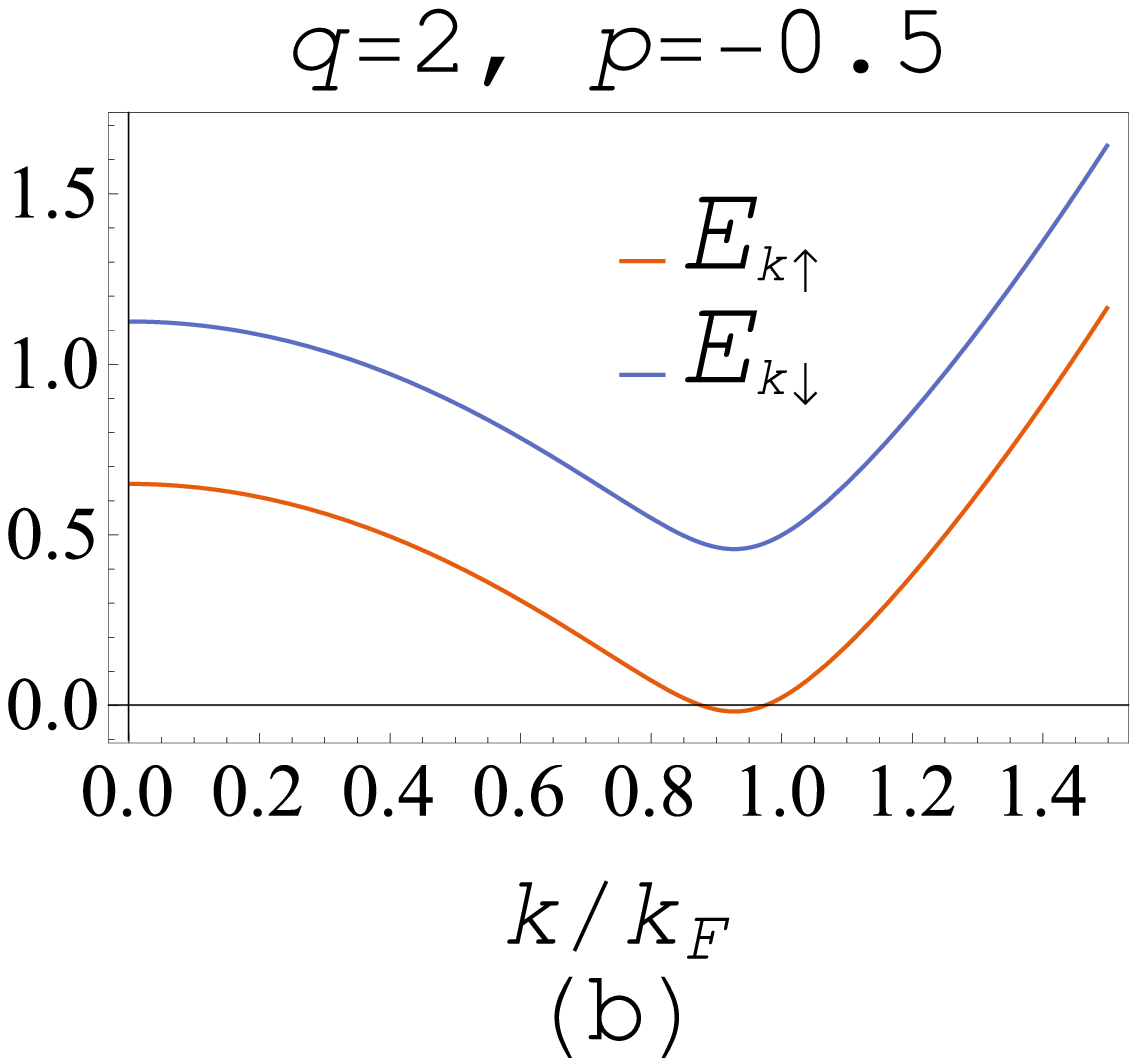}
 \caption{(Color online). Quasiparticle energy dispersions of the situations with $(q,p)=(0.5,0.5)$ and $(q,p)=(2,-0.5)$. $E_{\bk\uparrow}$ interchanges with $E_{\bk\downarrow}$ as $q\rightarrow \frac{1}{q}$ and $p\rightarrow -p$. }
 \label{fig.1}
\end{figure}

\section{Numerical Analysis}
Now we focus on the numerical solution of the model and try to visualize some interesting properties of it. For simplicity the temperature is restricted to be zero, and the coupling constant is chosen as $g=0.385/k^2_F$. We introduce the total number density $n=n_\uparrow+n_\downarrow$, number difference $\delta n=n_\uparrow-n_\downarrow$ and number polarization $p=\delta n /n$.
For convenience, in the following discussions the unit $k_F$ satisfying $n=k^3_F/(3\pi^2)$ is chosen as Fermi momentum of the un-deformed noninteracting Fermi gas with the same total particle of the $q$-deformed superconducting system under studying.

\begin{figure}[ht]
\centering
\includegraphics[width=2.4in, clip]{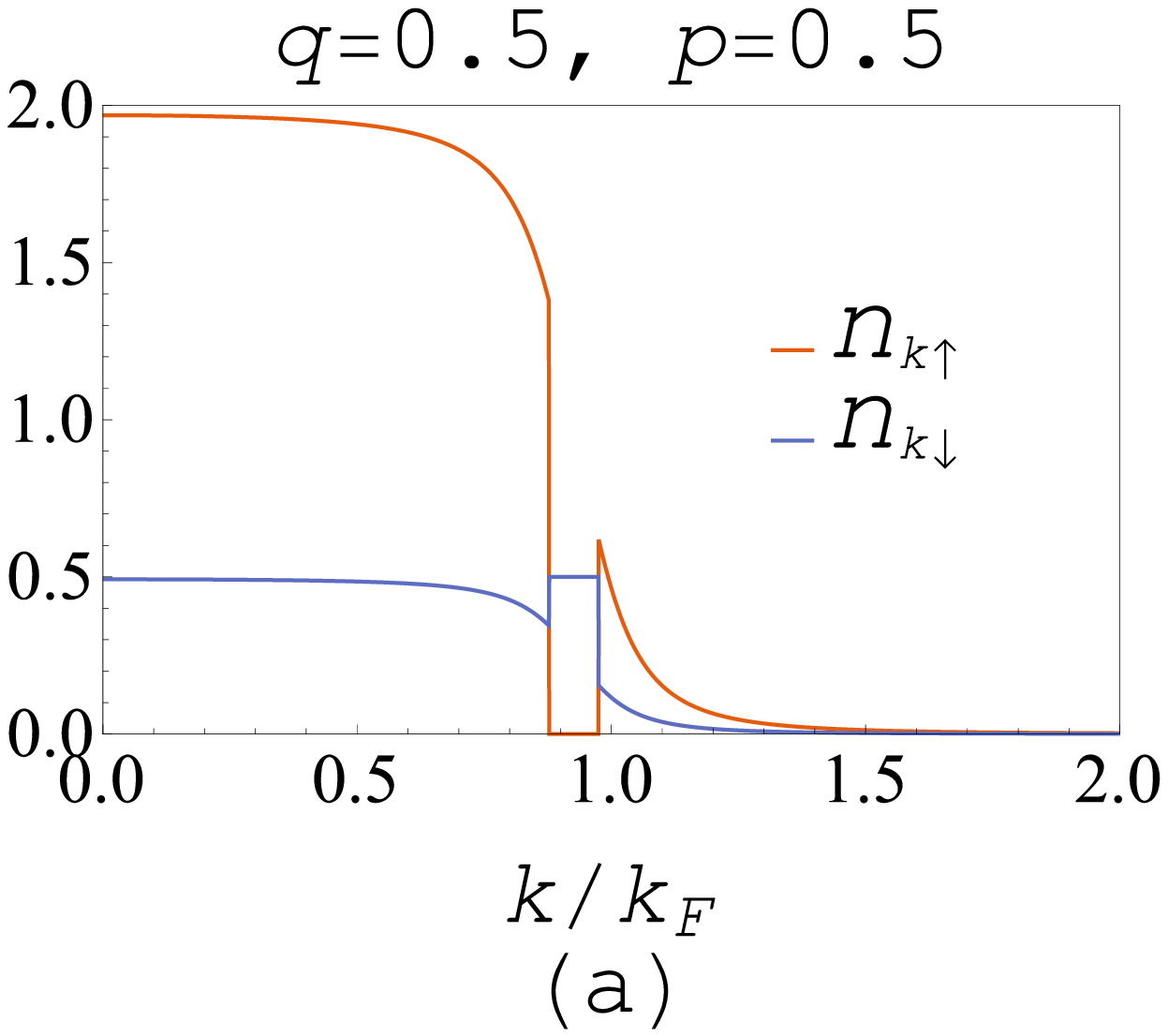}
\includegraphics[width=2.4in, clip]{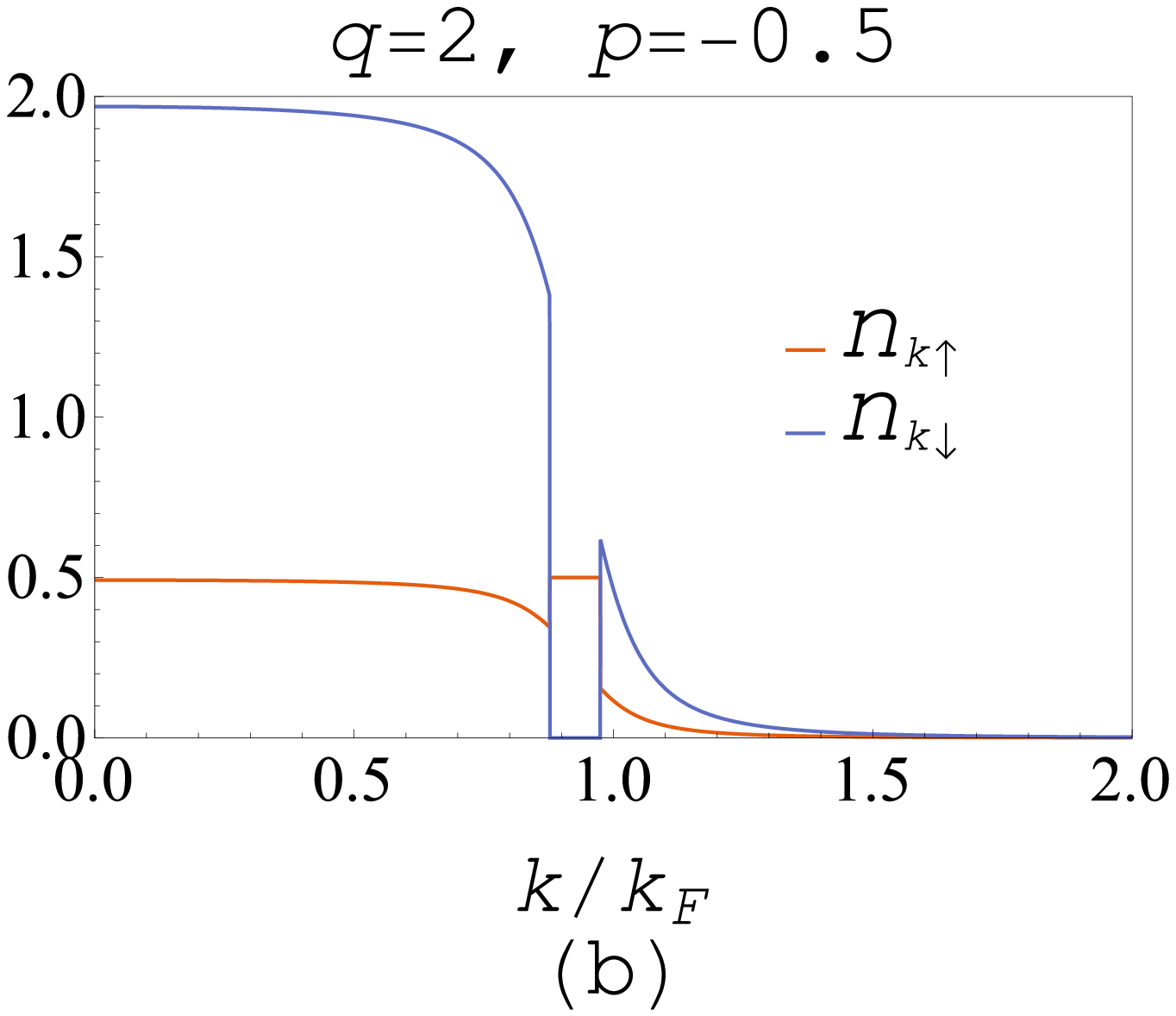}
 \caption{(Color online). Number density distributions as a function of $k$ for $(q,p)=(0.5,0.5)$ and $(q,p)=(2,-0.5)$. There is a regime where only one species (minority) exists. }
 \label{fig.2}
\end{figure}

As indicated from the equations of states, the system has a symmetry under the change $q\leftrightarrow \frac{1}{q}$, $p\leftrightarrow -p$ or $\uparrow \leftrightarrow\downarrow$, and other quantities changes simultaneously as $\mu\leftrightarrow\mu$, $h\leftrightarrow-h$ and $\Delta\leftrightarrow \frac{1}{\sqrt{q}}\Delta$. This symmetry can be explicitly observed by the quasiparticle energy dispersions, shown by Figure.\ref{fig.1}. Obviously in the two panels one can find that $E_{\bk\uparrow}\leftrightarrow E_{\bk\downarrow}$ as $q\rightarrow \frac{1}{q}$ and $p \rightarrow -p$. As in the $q=1$ situation, one of the energy dispersions becomes gapless in the region $[k_1,k_2]$ where \begin{align}&k_1=\textrm{max}(0, \sqrt{2m(\mu-\sqrt{(h+T\ln q)^2-\frac{\Delta^2}{q}})}),\notag\\ &k_2=\sqrt{2m(\mu+\sqrt{(h+T\ln q)^2-\frac{\Delta^2}{q}})},\end{align} i.e., $E_{\bk\uparrow}\le 0$ or $E_{\bk\downarrow}\le 0$ in $[k_1,k_2]$ and pairing in
momentum space is unfavorable in this regime. This can be seen most prominently
at $T=0$ when the Fermi distribution function becomes a step function, as presented by Fig.\ref{fig.2} (Obviously there is also a symmetry between panels (a) and (b) if $q\rightarrow \frac{1}{q}$ and $p\rightarrow -p$. The density distribution is similar to that of the Sarma phase\cite{Sarma} or breached pair state\cite{WilczekPRL03,DuanPRL06} since there is a regime that only one species exists. However, it is also different from the Sarma phase in some aspects. The first is that in the unpaired regimes, the existing particles belong to the minority particles, as indicated in panel (a), while in the ordinary Sarma phase, the situation is just the opposite. The second is that in the pairing regimes, i.e. when $k$ is below $k_1$ and above $k_2$, the particle numbers of the two species are unequal. Hence, it is reasonable to call this new ordered phase emerging from the $q$BCS model as the $q$-deformed Sarma phase, or $q$Sarma phase.
\begin{figure}[ht]
\centering
\includegraphics[width=2.4in, clip]{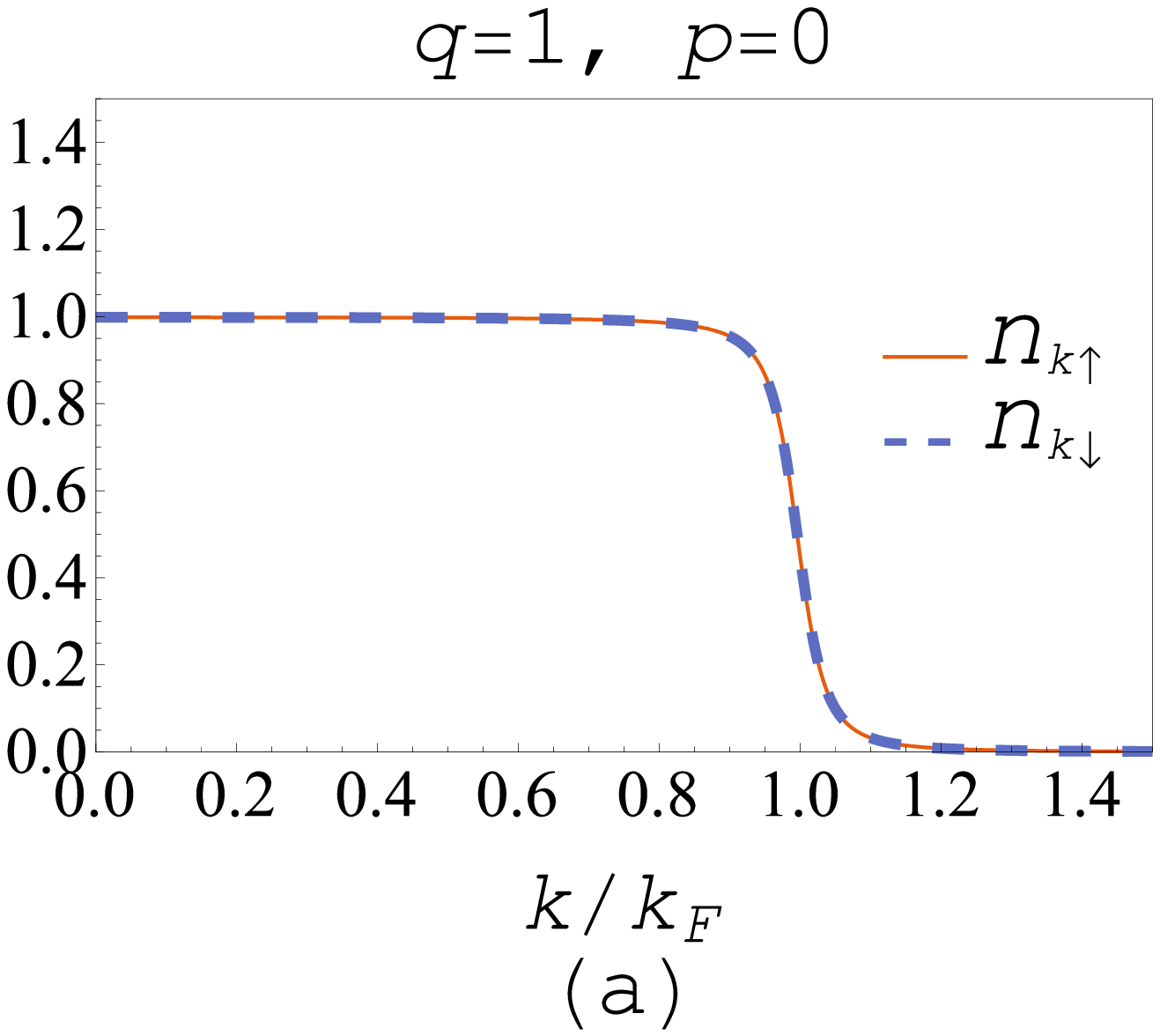}
\includegraphics[width=2.4in, clip]{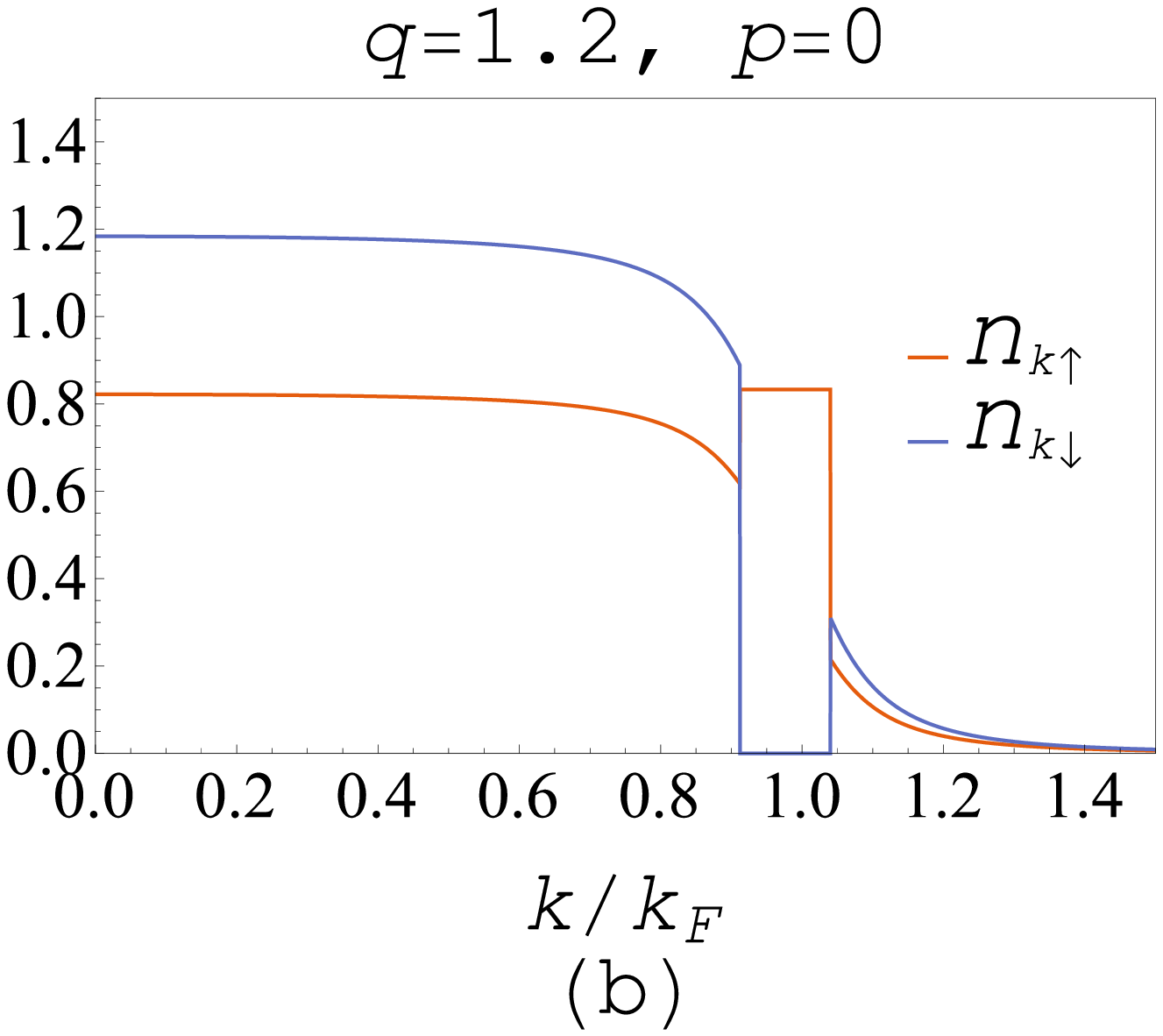}
 \caption{(Color online). Number density distributions as a function of $k$ for $q=1$, i.e. the ordinary BCS model, and $q=1.2$, i.e the $q$BCS model when $p=0$.}
 \label{fig.3}
\end{figure}

\begin{figure}[ht]
\centering
\includegraphics[width=3.3in, clip]{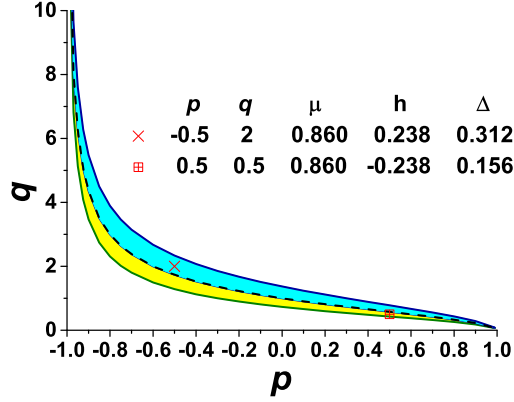}
 \caption{(Color online). Phase diagram on the $q$-$p$ plane. The $q$Sarma phase has two regions which are symmetric under the parameter transformation $q\rightarrow \frac{1}{q}$ and $p\rightarrow -p$. Here we also give the corresponding values of $\mu$, $h$ and $\Delta$ of two points which can be related by the parameter transformation. }
 \label{fig.4}
\end{figure}
There is a significant difference between the $q$BCS and the ordinary BCS model. Due to the deformation of different parameters, even when the particle number of each species are equal, the $q$BCS model still exhibits the breached pair superfluid. In Fig.\ref{fig.3} we present a diagram to show a comparison. In panel (a), $q=1$, i.e. there is no deformation and the model is simply the ordinary BCS model, obviously the pairing numbers are equal everywhere. While in panel (b), $q=1.2$, the number distribution is quite similar to those in Fig.\ref{fig.2} and the $q$Sarma phase still exists even when $p=0$.

In Fig.\ref{fig.4}, we summarize our observations in the form of a $q$-$p$ plane at $T=0$. The dashed and dotted lines indicate the boundaries between the normal ($q$-deformed non-interacting Fermi gas) and ordered ($q$Sarma) phases. The ordered regions can be divide into two parts, each part is the symmetric part of another under the parameter transformation $q\rightarrow \frac{1}{q}$ and $p\rightarrow -p$. The cyan region is characterized by $h>0$ while the yellow region is characterized by $h<0$.
The solid line denotes the dividing line of the two regions. It can be formally thought of as where $h=0$ though numerically no solution in fact exists there. As an example of the symmetry, we also mark two points in the two regions of the ordered phase respectively, and explicitly give the values of the corresponding $\mu$, $h$ and $\Delta$. One can clearly see that under the parameter transformation $q\rightarrow \frac{1}{q}$ and $p\rightarrow -p$, these quantities change as $\mu\rightarrow \mu$, $h\rightarrow-h$ and $\Delta\rightarrow \frac{1}{\sqrt{q}}\Delta$.

The phase diagram also presents an interesting feature that as $q\rightarrow 0$ or $+\infty$ the ordered phase only appears when $p\rightarrow 1$ or $-1$ respectively. Here we provide an intuitive explanation for the situation with $q\rightarrow 0$ and $p\rightarrow 1$, the other one can be understood accordingly by noting the $q\leftrightarrow \frac{1}{q}$ and $p\leftrightarrow -p$ symmetry of the system. It is always more favorable for the two species of $q$-fermions to pair up if the Fermi surfaces of them are closer to each other\cite{LYD06}. Since the deformation parameters of $q$-fermions with up and down spins are $q$ and $\frac{1}{q}$ respectively, then indicated by Eq.(\ref{nfm}) the Fermi momentum of each species are
\begin{align}
k_{F\uparrow}=\sqrt[3]{6q\pi^2n_\uparrow},\quad k_{F\downarrow}=\sqrt[3]{\frac{6}{q}\pi^2n_\downarrow},
\end{align}
if the there is no interaction between the fermions.
As $p\rightarrow 1$, it is easy to find $n_\uparrow\gg n_\downarrow$, hence $k_{F\uparrow}\approx k_{F\downarrow}$ only when $0<q\ll 1$, i.e. the ordered phase is only favored at this region. As an example, we choose a point lying in the region of the $q$Sarma phase from Fig.\ref{fig.4}, of which the parameters are $p\approx0.99$, $q\approx 0.074$, the corresponding Fermi momenta are found to be $k_{F\uparrow}\approx 0.147$, $k_{F\downarrow}\approx 0.149$. They are indeed close to each other and the pairing is favored here.

\section{Thermodynamics}
We now generalize our previous discussions about the thermodynamics for $q$-deformed non-interacting Fermi gas to the $q$BCS model. The central task is to evaluate the partition function given by Eq.(\ref{gpf}) where $H$ is the hamiltonian (\ref{HqBCS}) of the $q$BCS model. The trace can be taken in the Fock space of the quasiparticles, i.e. $Z=\sum_n\langle n|\me^{-\beta H}|n\rangle$ where $n$ is number of quasiparticles, since the Hamiltonian can be recognized as being diagonalized in this space just as the ordinary BCS model.
We recognize that the quasiparticle $\alpha_\bk$ has the energy dispersion $E_{\bk\uparrow}-\frac{\ln q}{\beta}$, and $\beta_\bk$ has the energy dispersion $E_{\bk\downarrow}+\frac{\ln q}{\beta}$, hence we have
 \begin{align}
Z
&=\sum_{n_{\alpha_1}+\cdots+n_{\alpha_\infty}+n_{\beta_1}+\cdots+n_{\beta_\infty}=n}
\langle n_{\alpha_1},\cdots ,n_{\alpha_\infty},n_{\beta_1},\cdots ,n_{\beta_\infty}|\me^{-\beta H}|n_{\alpha_1},\cdots , n_{\alpha_\infty},n_{\beta_1},\cdots ,n_{\beta_\infty}\rangle\notag\\
&=\me^{-\beta[\sum_{\bk}(\xi_{\bk}-E_{\bk})+\frac{|\Delta|^2}{qg}]}\sum_{n_{\alpha_1}}\langle n_{\alpha_1}|\me^{-\beta (E_{1\uparrow}-\frac{\ln q}{\beta}) n_{\alpha_1}}|n_{\alpha_1}\rangle\cdots\sum_{n_{\alpha_\infty}}\langle  n_{\alpha_\infty}|\me^{-\beta (E_{\infty\uparrow}-\frac{\ln q}{\beta}) n_{\alpha_\infty}}|n_{\alpha_\infty}\rangle  \nonumber \\
&\times \sum_{n_{\beta_1}}\langle  n_{\beta_1}|\me^{-\beta (E_{1\downarrow}+\frac{\ln q}{\beta}) n_{\beta_1}}|n_{\beta_1}\rangle\cdots\sum_{n_{\beta_\infty}}\langle n_{\beta_\infty}|\me^{-\beta (E_{\infty\downarrow}+\frac{\ln q}{\beta}) n_{\beta_\infty}}|n_{\beta_\infty}\rangle\notag\\
&=\me^{-\beta[\sum_{\bk} (\xi_{\bk}-E_{\bk})+\frac{|\Delta|^2}{qg}]}\big[\prod^{\infty}_\bk \sum^{1}_{n_\bk=0}\me^{-\beta (E_{\bk\uparrow}-\frac{\ln q}{\beta})n_{\bk}}\big]\big[\prod^{\infty}_\bk \sum^{1}_{n_\bk=0}\me^{-\beta (E_{\bk\downarrow}+\frac{\ln q}{\beta})n_{\bk}}\big]\notag\\
&=\me^{-\beta\frac{|\Delta|^2}{qg}}\prod^{\infty}_{\bk} \me^{-\beta (\xi_{\bk}-E_{\bk})}(1+\me^{-\beta (E_{\bk\uparrow}-\frac{\ln q}{\beta})})(1+\me^{-\beta (E_{\bk\downarrow}+\frac{\ln q}{\beta})}),
\end{align}
where $\alpha_i\equiv\alpha_{\bk_i}$, $\beta_i\equiv\beta_{\bk_i}$, and $E_{i\uparrow,\downarrow}\equiv E_{\bk_i\uparrow,\downarrow}$. Then the thermodynamic potential is given by
\begin{align}\label{qOBCS}
\Omega= -\frac{1}{\beta}\ln Z = \frac{|\Delta|^2}{qg}+\sum_{\bk}(\xi_{\bk}-E_{\bk})-\frac{1}{\beta}\sum_{\bk}\ln(1+\me^{-\beta (E_{\bk\uparrow}-\frac{\ln q}{\beta})})-\frac{1}{\beta}\sum_{\bk}\ln(1+\me^{-\beta (E_{\bk\downarrow}+\frac{\ln q}{\beta})}),
\end{align}
and the entropy is evaluated as
\begin{align}
S&=-\frac{\partial \Omega}{\partial T} \nonumber \\
&=-\sum_\bk\Big[ f(E_{\bk\uparrow}-\frac{\ln q}{\beta})\ln\frac{1}{q}f(E_{\bk\uparrow}-\frac{\ln q}{\beta})+
f(-E_{\bk\uparrow}+\frac{\ln q}{\beta})\ln f(-E_{\bk\uparrow}+\frac{\ln q}{\beta})\Big]\notag\\
&\quad-\sum_\bk\Big[ f(E_{\bk\downarrow}+\frac{\ln q}{\beta})\ln qf(E_{\bk\downarrow}+\frac{\ln q}{\beta})+
f(-E_{\bk\downarrow}-\frac{\ln q}{\beta})\ln f(-E_{\bk\downarrow}-\frac{\ln q}{\beta})\Big].
\end{align}
The evaluation of the total energy needs careful treatments. We can not directly apply either Eq.(\ref{qE}) or Eq.(\ref{qEOSN}) since the deformation parameters of the two species are different. Moreover, the thermodynamical potential can not be separated into two parts associated with spin-up and spin-down fermions respectively since they form Cooper pairs and become condensed. However, since the system becomes a non-interacting $q$-Fermi gas when $\Delta\rightarrow 0$,
then the thermodynamical potential can be split into $\Omega=\Omega_\uparrow+\Omega_\downarrow$ in that limit.
Hence it is natural to require the total energy satisfy the property
\begin{align}\label{qE2}
E=E_\uparrow+E_\downarrow\xrightarrow{\Delta\rightarrow 0}\frac{1}{q}(\Omega_\uparrow+TS_\uparrow)+\mu_\uparrow n_\uparrow+q(\Omega_\downarrow+TS_\downarrow)+\mu_\downarrow n_\downarrow.
\end{align}
Hence we generalize the formalism given in Ref.\cite{ChenQJthesis} to give
\begin{align}\label{qEBCS0}
E=\sum_K\Big[\big(\epsilon_\bk+\frac{1}{2}\Sigma_\uparrow(K)\big)\frac{1}{q}G_\uparrow(K)+\big(\epsilon_\bk+\frac{1}{2}\Sigma_\downarrow(K)\big)qG_\downarrow(K)\Big].
\end{align}
It can be easily shown that this identity reduces to Eq.(\ref{qE2}) if $\Delta\rightarrow 0$, and reduces to the total energy of the ordinary BCS theory if $q=1$. By a straightforward calculation, the expression of the total energy is
\begin{align}
&E=\sum_{\bk} \Bigg\{(\frac{q}{2}+\frac{1}{2q})(\xi_\bk-E_\bk)-(h+\frac{\ln q}{\beta})(\frac{1}{2q}-\frac{q}{2})(1-\frac{\xi_\bk}{E_\bk})+(\frac{1}{2q}-\frac{q}{2})\xi_\bk\Big[f(E_{\bk\uparrow}-\frac{\ln q}{\beta})-f(E_{\bk\downarrow}+\frac{\ln q}{\beta})\Big]\notag\\& +(\frac{1}{2q}+\frac{q}{2})\Big[(E_{\bk\uparrow}-\frac{\ln q}{\beta})f(E_{\bk\uparrow}-\frac{\ln q}{\beta})+(E_{\bk\downarrow}+\frac{\ln q}{\beta})f(E_{\bk\downarrow}+\frac{\ln q}{\beta})\Big]
-(h+\frac{\ln q}{\beta})(\frac{1}{2q}-\frac{q}{2})\frac{\xi_\bk}{E_\bk}\Big[f(E_{\bk\uparrow}-\frac{\ln q}{\beta})+f(E_{\bk\downarrow}+\frac{\ln q}{\beta})\Big]\Bigg\}\notag\\& -(1+\frac{1}{q^2})\frac{|\Delta|^2}{2g}+\mu_\uparrow n_\uparrow+\mu_\downarrow n_\downarrow.
\end{align}
Here we emphasize that there is not a simple relation similar to Eq.(\ref{qEOSN}) exists, only when $\Delta\rightarrow 0$ we have the relation (\ref{qE2}).

The equations of states can also be obtained by differentiating the thermodynamical potential
\begin{align}
n_{\uparrow}&=-\frac{\partial \Omega}{q\partial \mu_{\uparrow}} ,\notag\\
n_{\downarrow}&=-\frac{q\partial \Omega}{\partial \mu_{\downarrow}},\notag\\
\frac{\partial \Omega}{\partial \Delta}&=0.
\end{align}
Here we emphasize that the deformation parameter $\frac{1}{q}$ must be included in the first term of $\Omega$ (see the first term of the second line of Eq.(\ref{qOBCS})) to give the correct gap equation.

\section{Conclusion}
In summary, we construct the finite temperature formalism for the $q$-deformed many-fermion system since applications of the $q$-deformed statistics has found  utilizations in several fields of physics. We first studied the $q$-deformed non-interacting Fermi gas. Interestingly, the Matsubara frequency of the finite temperature Green's function has an extra imaginary part. This formalism is further generalized to the well-known interacting Fermi system, the BCS model. Importantly, to ensure that the dynamical equations of the Green's functions to be closed, we find that the deformation parameters of one species must be the reciprocal of the other. We obtained the equations of states, and found that this model has a symmetry in the parameter space and exhibits some interesting properties. Its ordered phase is a generalization of the Sarma phase, i.e., the $q$Sarma phase. Finally, we presented the revised thermodynamic relations of this model in the presence of deformation parameter.

\textit{Acknowledgment}:  H. G. thanks the support from the National
Natural Science Foundation of China (Grant No.
11674051).

\appendix
\section{Properties of the Green's function of fermionic noninteracting $q$-gas}\label{appa1}
To study the periodicity property of the Green's function, we choose $x=(0,\bx)$, $x'=(\tau,\bx')$ with $0\leq \tau\leq \beta$, by the definition of the Green's function we have
	\begin{align}
	G(x,x')
	=q\frac{1}{Z}\textrm{Tr}\left(\me^{-\beta H}\psi^\dagger(x')\psi(x)\right)=q\frac{1}{Z}\textrm{Tr}\left(\me^{-\beta H}\me^{\beta H}\psi(0,\bx)\me^{-\beta H}\psi^\dagger(\tau,\bx')\right)=q\langle\psi(\beta,\bx)\psi^\dagger(\tau,\bx')\rangle\theta(\beta-\tau)=-qG(\beta\bx,\tau\bx').
	\end{align}
Here we assume the system is homogenous and hence the Green's function has a spacetime translational symmetry, i.e., $G(x,x')=G(x-x')$. Therefore we have
\begin{align}
G(-\tau,\bx-\bx')=-qG(-\tau+\beta,\bx-\bx').\label{p1}
\end{align}
It's convenient to separate the Fourier transformation form (\ref{FTG}) into two parts.
	\begin{eqnarray}
	G(\omega_n,\bx)=\frac{1}{2}\int^0_{-\beta}d\tau \me^{\mi\omega_n\tau}G(\tau,\bx)+\frac{1}{2}\int^{\beta}_0 d\tau \me^{\mi\omega_n\tau}G(\tau,\bx).
	\end{eqnarray}
Applying the boundary condition (\ref{p1}) and changing the variable, we get
	\begin{align}
	G(\omega_n,\bx)
	 =\frac{1}{2}(-q\me^{-\mi\omega_n\beta}+1)\int^\beta_0 d\tau \me^{\mi\omega_n\tau}G(\tau,\bx).
	\end{align}
We hope that $1-q\me^{-\mi\omega_n\beta}=1-(-1)^n$ and the Fourier transformation can be expressed as
\begin{align}
G(\omega_n,\bx)=\int^\beta_0 d\tau \me^{\mi\omega_n\tau}G(\tau,\bx).
\end{align}
Therefore the condition $q\me^{-\mi\omega_n\beta}=(-1)^n$ leads to the fact that the Matsubara frequency $\omega_n$ is a complex number	
\begin{eqnarray}\label{mf1}
	\omega_n=\frac{(2n+1)\pi}{\beta}-\mi\frac{\ln q}{\beta}.
	\end{eqnarray}

The Green's function can be evaluated by studying the equation of motion of it. We have
\begin{align}
	\frac{\partial G(x,x')}{\partial		\tau}&=-\langle\psi(x)\psi^\dagger(x')\rangle\delta(\tau-\tau')-q\langle\psi^\dagger(x')\psi(x)\rangle\delta(\tau-\tau')-\left(\langle\frac{\partial \psi(x)}{\partial \tau} \psi^\dagger(x')\rangle\theta(\tau-\tau')-q\langle\psi^\dagger(x')\frac{\partial \psi(x)}{\partial \tau}\rangle\theta(\tau'-\tau)\right)\notag\\
&=-\langle\big[\psi(x)\psi^\dagger(x')+q\psi^\dagger(x')\psi(x)\big]\rangle\delta(\tau-\tau')-\langle T_\tau\Big[ \frac{\partial \psi(x)}{\partial \tau}\psi^\dagger(x')\Big]\rangle.
	\end{align}
By applying the cyclic property of the trace, the first term is evaluated as
\begin{align}
& \langle\big[\psi(x)\psi^\dagger(x')+q\psi^\dagger(x')\psi(x)\big]\rangle\delta(\tau-\tau')\notag\\
&=\langle \big[\me^{H\tau}\psi(\bx)\me^{-H\tau} \me^{H \tau'} \psi^\dagger(\bx') \me^{-H\tau'}+q\me^{H\tau'}\psi^\dagger(\bx') \me^{-H\tau'}\me^{H\tau}\psi(\bx)\me^{-H\tau}\delta(\tau-\tau')\big]\rangle\nonumber\\
&=\delta(\bx-\bx')\delta(\tau-\tau').
\end{align}
To evaluate the second term, we need to apply the equation of motion for the field operator. Given the definition of the Heisenberg operator $\psi(x)=\me^{H\tau}\psi(\bx)\me^{-H\tau}$, we have
	\begin{eqnarray}
	\frac{\partial \psi(x)}{\partial \tau}=\me^{H\tau}[H,\psi(\bx)]\me^{-H\tau}=-\big(\frac{-\nabla^2}{2m}-\mu\big)\psi(x).
	\end{eqnarray}
Therefore we get $\langle T_\tau \big[\frac{\partial \psi(x)}{\partial \tau}\psi^\dagger(x')\big]\rangle=(\frac{-\nabla^2}{2m}-\mu)G(x,x')$.
The Green's function then satisfy the differential equation
	\begin{eqnarray}\label{eom0}
	\left(-\frac{\partial }{\partial \tau}-\frac{-\nabla^2}{2m}+\mu\right)G(x,x')&=&\delta(\bx-\bx')\delta(\tau-\tau').
	\end{eqnarray}

\section{Properties of the Green's function of $q$BCS theory}\label{appb1}
Applying the equations of motion (\ref{eomf}) for field operators, we get
\begin{align}\label{eomu}
&\frac{\partial G_{\uparrow}(x,x')}{\partial \tau}=-\langle\psi_{\uparrow}(x)\psi^{\dag}_{\uparrow}(x')\rangle\delta(\tau-\tau')-q\langle\psi^{\dag}_{\uparrow}(x')\psi_{\uparrow}(x)\rangle\delta(\tau'-\tau)
-\langle \frac{\partial \psi_{\uparrow}(x)}{\partial \tau}\psi^{\dag}_{\uparrow}(x')\rangle\theta(\tau-\tau')+q\langle\psi^{\dag}_{\uparrow}(x')\frac{\partial \psi_{\uparrow}(x)}{\partial \tau}\rangle\theta(\tau'-\tau) \nonumber \\
&=-\delta(\bx-\bx')\delta(\tau-\tau')-\langle\Big[ -(\frac{(-\mi\nabla)^2}{2m}-\mu_{\uparrow})\psi_{\uparrow}(x)\psi^{\dag}_{\uparrow}(x')+q^{-1}\Delta(\bx)\psi^\dag_{\downarrow}(x)\psi^{\dag}_{\uparrow}(x')\Big]\rangle\theta(\tau-\tau') \notag\\&+q\langle\Big[-(\frac{(-\mi\nabla)^2}{2m}-\mu_{\uparrow})\psi^{\dag}_{\uparrow}(x')\psi_{\uparrow}(x)+q^{-1}\Delta(\bx)\psi^{\dag}_{\uparrow}(x')\psi^\dag_{\downarrow}(x)\Big]\rangle\theta(\tau'-\tau) \nonumber \\
&=-\delta(\bx-\bx')\delta(\tau-\tau')-(\frac{(-\mi\nabla)^2}{2m}-\mu_{\uparrow})G_{\uparrow}(x,x')+q^{-1}\Delta(\bx)F^{\dag}_{\uparrow\downarrow}(x,x'). \end{align}
One can find that the introduction of the symmetric parameters $q$ and $\frac{1}{q}$ in the algebra (\ref{qBCSa1}) is crucial to lead to the last line of Eq.(\ref{eomu}) so as to form a closed set of differential equations. Similarly we have
\begin{align}
&\frac{\partial F^{\dag}_{\uparrow\downarrow}(x,x')}{\partial \tau}=-\langle\psi^{\dag}_{\downarrow}(x)\psi^{\dag}_{\uparrow}(x')\rangle\delta(\tau-\tau')-q\langle\psi^{\dag}_{\uparrow}(x')\psi^{\dag}_{\downarrow}(x)\rangle\delta(\tau'-\tau)    -\langle\frac{\partial \psi^{\dag}_{\downarrow}(x)}{\partial \tau}\psi^{\dag}_{\uparrow}(x')\rangle\theta(\tau-\tau')+q\langle\psi^{\dag}_{\uparrow}(x')\frac{\partial \psi^{\dag}_{\downarrow}(x)}{\partial \tau}\rangle\theta(\tau'-\tau) \nonumber \\
&=-\langle\Big[(\frac{(-\mi\nabla)^2}{2m}-\mu_{\downarrow})\psi^{\dag}_{\downarrow}(x)\psi^{\dag}_{\uparrow}(x')+\Delta^{*}(\bx)\psi_{\uparrow}(x)\psi^{\dag}_{\uparrow}(x')\Big]\rangle\theta(\tau-\tau') +q\langle\Big[(\frac{(-\mi\nabla)^2}{2m}-\mu_{\downarrow})\psi^{\dag}_{\uparrow}(x')\psi^{\dag}_{\downarrow}(x)+\Delta^{*}(\bx)\psi^{\dag}_{\uparrow}(x')\psi_{\uparrow}(x)\Big]\rangle\theta(\tau'-\tau) \nonumber \\
&=(\frac{(-\mi\nabla)^2}{2m}-\mu_{\downarrow})F^{\dag}_{\uparrow\downarrow}(x,x')+\Delta^{*}(\bx)G_{\uparrow}(x,x').
\end{align}
Rearranging the terms, we obtain the equations of motion for the Green's function and anomalous Green's function
\begin{align}
&(-\frac{\partial}{\partial \tau}-\frac{(-\mi\nabla)^2}{2m}+\mu_{\uparrow})G_{\uparrow}(x,x')+q^{-1}\Delta(\bx)F^{\dag}_{\uparrow\downarrow}(x,x')=\delta(\bx-\bx')\delta(\tau-\tau'),  \notag\\
&(\frac{\partial}{\partial \tau}-\frac{(-\mi\nabla)^2}{2m}+\mu_{\downarrow})F^{\dag}_{\uparrow\downarrow}(x,x')=\Delta^{*}(\bx)G_{\uparrow}(x,x').
\end{align}
In almost all situations of interest, the system has both temporal and spatial translation symmetries, hence we have $G_{\sigma}(x,x')=G_{\sigma}(x-x')$ and $F_{\sigma\bar{\sigma}}(x,x')=F_{\sigma\bar{\sigma}}(x-x')$. By implementing Fourier transformations, we get
\begin{align}
&(\mi\omega_n+\frac{\ln q}{\beta}-\xi_{\bk\uparrow})G_{\uparrow}(\mi\omega_n,\bk)+q^{-1}\Delta F^{\dag}_{\uparrow\downarrow}(\mi\omega_n,\bk)=1,\notag  \\
&(-\mi\omega_n-\frac{\ln q}{\beta}-\xi_{\bk\downarrow})F^{\dag}_{\uparrow\downarrow}(\mi\omega_n,\bk)-\Delta^{*}G_{\uparrow}(\mi\omega_n,\bk)=0,
\end{align}
which can be readily solved as
\begin{align}
G_{\uparrow}(\mi\omega_n,\bk)&=\frac{\mi\omega_n+\frac{\ln q}{\beta}+\xi_{\bk\downarrow}}{(\mi\omega_n+\frac{\ln q}{\beta}-\xi_{\bk\uparrow})(\mi\omega_n+\frac{\ln q}{\beta}+\xi_{\bk\downarrow})-q^{-1}|\Delta|^2}, \nonumber \\
F^{\dag}_{\uparrow\downarrow}(\mi\omega_n,\bk)&=\frac{-\Delta^{*}}{(\mi\omega_n+\frac{\ln q}{\beta}-\xi_{\bk\uparrow})(\mi\omega_n+\frac{\ln q}{\beta}+\xi_{\bk\downarrow})-q^{-1}|\Delta|^2}.
\end{align}
They can be further simplified as
\begin{align} \label{eud2}
& G_{\uparrow}(\mi\omega_n,\bk)
=\frac{\mi\omega_n+\frac{\ln q}{\beta}+\xi_{\bk}+h}{(\mi\omega_n+\frac{\ln q}{\beta}+h)^2-E_{\bk}^2}=\frac{|u_{\bk}|^2}{\mi\omega_n+\frac{\ln q}{\beta}+h-E_{\bk}}+\frac{|v_{\bk}|^2}{\mi\omega_n+\frac{\ln q}{\beta}+h+E_{\bk}},  \\
&F^{\dag}_{\uparrow\downarrow}(\mi\omega_n,\bk)
=\frac{-\Delta^{*}}{(\mi\omega_n+\frac{\ln q}{\beta}+h)^2-E_{\bk}^2}=-\sqrt{q} u_{\bk}^{*}v_{\bk}\Big[\frac{1}{\mi\omega_n+\frac{\ln q}{\beta}+h-E_{\bk}}-\frac{1}{\mi\omega_n+\frac{\ln q}{\beta}+h+E_{\bk}}\Big],
\end{align}
where we have plugged in the relations
\begin{align}
|u_{\bk}|^2+|v_{\bk}|^2=1, |u_{\bk}|^2-|v_{\bk}|^2=\frac{\xi_{\bk}}{E_{\bk}}, u_{\bk}v^{*}_{\bk}=\frac{\Delta}{2\sqrt{q} E_{\bk}}, \xi_{\bk\uparrow}=\xi_{\bk}-h, \xi_{\bk\downarrow}=\xi_{\bk}+h, E_{\bk}=\sqrt{\xi_{\bk}^2+q^{-1}|\Delta|^2}.
\end{align}
The equations of motion for the Green's function $G_{\downarrow}$ and $F^\dagger_{\downarrow\uparrow}$ can deduced from Eqs.(\ref{eomf}) by
exactly the same way. And we finally get
\begin{align}
&(-\frac{\partial}{\partial \tau}-\frac{(-\mi\nabla)^2}{2m}+\mu_{\downarrow})G_{\downarrow}(x,x')-\Delta(\bx)F^{\dag}_{\downarrow\uparrow}(x,x')=\delta(\bx-\bx')\delta(\tau-\tau'), \notag \\
&(\frac{\partial}{\partial \tau}-\frac{(-\mi\nabla)^2}{2m}+\mu_{\uparrow})F^{\dag}_{\downarrow\uparrow}(x,x')=-q^{-1}\Delta^{*}(\bx)G_{\downarrow}(x,x').
\end{align}
By implementing the Fourier transformation, these equations become
\begin{align}
&(\mi\omega_n-\frac{\ln q}{\beta}-\xi_{\bk\downarrow})G_{\downarrow}(\mi\omega_n,\bk)-\Delta F^{\dag}_{\downarrow\uparrow}(\mi\omega_n,\bk)=1,  \notag\\
&(-\mi\omega_n+\frac{\ln q}{\beta}-\xi_{\bk\uparrow})F^{\dag}_{\downarrow\uparrow}(\mi\omega_n,\bk)+q^{-1}\Delta^{*}G_{\downarrow}(\mi\omega_n,\bk)=0,
\end{align}
which can be readily solved as
\begin{align}
 G_{\downarrow}(\mi\omega_n,\bk)&=\frac{\mi\omega_n-\frac{\ln q}{\beta}+\xi_{\bk\uparrow}}{(\mi\omega_n-\frac{\ln q}{\beta}-\xi_{\bk\downarrow})(\mi\omega_n-\frac{\ln q}{\beta}+\xi_{\bk\uparrow})-q^{-1}|\Delta|^2}, \nonumber \\
 \quad F^{\dag}_{\downarrow\uparrow}(\mi\omega_n,\bk)&=\frac{q^{-1}\Delta^{*}}{(\mi\omega_n-\frac{\ln q}{\beta}-\xi_{\bk\downarrow})(\mi\omega_n-\frac{\ln q}{\beta}+\xi_{\bk\uparrow})-q^{-1}|\Delta|^2}.
\end{align}
They can be further expressed as
\begin{align} \label{edu2}
 &G_{\downarrow}(\mi\omega_n,\bk)
=\frac{\mi\omega_n-\frac{\ln q}{\beta}+\xi_{\bk}-h}{(\mi\omega_n-\frac{\ln q}{\beta}-h)^2-E_{\bk}^2}=\frac{|u_{\bk}|^2}{\mi\omega_n-\frac{\ln q}{\beta}-h-E_{\bk}}+\frac{|v_{\bk}|^2}{\mi\omega_n-\frac{\ln q}{\beta}-h+E_{\bk}},  \\
& F^{\dag}_{\downarrow\uparrow}(\mi\omega_n,\bk)
=\frac{q^{-1}\Delta^{*}}{(\mi\omega_n-\frac{\ln q}{\beta}-h)^2-E_{\bk}^2}=\frac{u_{\bk}^{*}v_{\bk}}{\sqrt{q}}
\Big[\frac{1}{\mi\omega_n-\frac{\ln q}{\beta}-h-E_{\bk}}-\frac{1}{\mi\omega_n-\frac{\ln q}{\beta}-h+E_{\bk}}\Big].
\end{align}

\end{document}